\documentclass[american,aps,prx,reprint,superscriptaddress,nofootinbib,longbibliography,floatfix]{revtex4-2}

\usepackage{dcolumn}    
\usepackage{bm} 
\usepackage{graphicx}
\usepackage{lipsum}
\usepackage{amsmath}  
\usepackage{amsthm}
\usepackage{enumitem}
\usepackage{latexsym}
\usepackage{amsfonts} 
\usepackage{physics}
\usepackage{amssymb}
\usepackage{array}      
\usepackage{epsfig}
\usepackage{braket} 
\usepackage{bbold}
\usepackage{color}
\usepackage{comment}
\usepackage{soul}
\usepackage{xcolor}
\usepackage[normalem]{ulem}
\usepackage{tikz}
\usetikzlibrary{quantikz}
\usepackage[
    colorlinks=true,
    breaklinks=true,
    linkcolor=blue,
    urlcolor=blue,
    citecolor=blue,
    pdfusetitle
]{hyperref}
\usepackage{xurl}

\usetikzlibrary{decorations.pathreplacing}

\tikzset{
noise/.style={
    fill=orange!40,
    rounded corners=2pt, 
    minimum width=0.6cm,
    minimum height=0.7cm
},
recovery/.style={
    fill=blue!20,
    rounded corners=2pt,
    minimum width=0.6cm,
    minimum height=0.7cm
}
}

\theoremstyle{definition}
\newtheorem{definition}{Definition}
\newtheorem{proposition}{Proposition}
\newtheorem{theorem}{Theorem}
\newtheorem{lemma}{Lemma}
\newtheorem{corollary}{Corollary}

\DeclareMathOperator{\idChan}{id}

\def\mE{\mathcal{E}}

\def\mN{\mathcal{N}}
\def\mM{\mathcal{M}}

\def\mD{\mathcal{D}}

\def\mU{\mathcal{U}}

\DeclareMathOperator{\Herm}{Herm}

\begin{document}

\begin{abstract}

Noise is omnipresent in quantum systems, and thus, approximately reversing this noise is a central challenge towards quantum technologies. The Petz map provides a physically admissible recovery channel, but its performance depends critically on the choice of reference state. Here, we introduce and investigate the iterative Petz recovery (IPR) protocol, an adaptive multicopy strategy that begins from an unbiased reference state and uses each recovered output to construct the Petz map for the next iteration. We establish the stability of the protocol and develop a tomography-assisted finite-copy implementation with explicit control of error propagation and fidelity loss. For qubit depolarizing noise, we derive an exact recurrence for the fidelity of every pure input. For single-axis Pauli noise, we derive an exact recurrence for the six-state Pauli average and compare it numerically with the Haar average. We further benchmark finite-copy IPR against measure-and-prepare strategies based on local and collective measurements, revealing the associated finite-resource tradeoffs. 
\end{abstract}

\title{Iterative Petz Protocols for Quantum State Recovery}

\author{Lea Lautenbacher}
\email{lea.lautenbacher@uni-ulm.de}
\affiliation{Institute of Theoretical Physics, Ulm University, Albert-Einstein-Allee 11 89081, Ulm, Germany}

\author{Felix Ahnefeld}
\email{felix.ahnefeld@uni-ulm.de}
\affiliation{Institute of Theoretical Physics, Ulm University, Albert-Einstein-Allee 11 89081, Ulm, Germany}

\author{Susana F. Huelga}
\affiliation{Institute of Theoretical Physics, Ulm University, Albert-Einstein-Allee 11 89081, Ulm, Germany}
\affiliation{Center for Integrated Quantum Science and Technology (IQST), 89081 Ulm, Germany }

\author{Martin B. Plenio}
\affiliation{Institute of Theoretical Physics, Ulm University, Albert-Einstein-Allee 11 89081, Ulm, Germany}
\affiliation{Center for Integrated Quantum Science and Technology (IQST), 89081 Ulm, Germany }

\maketitle

\section{Introduction}
Quantum information processing tasks are fundamentally limited by noise introduced by quantum channels. Therefore, mitigating or even reversing the action of certain quantum channels is of central interest in quantum technologies. However, generic noise channels do not admit an exact completely positive and trace-preserving (CPTP) inverse \cite{Kadison1951,Choi1975}, which leads to the question of whether quantum channels can be (approximately) reversed when acting on a restricted set of input states. 

Among the various strategies for reversing quantum noise, the \textit{Petz recovery map} has emerged as a mathematically elegant and operationally meaningful tool, as it is directly connected to the saturation of the data processing inequality for the quantum relative entropy \cite{Petz1986,Petz1988}. Over the past decade, the Petz map has also gained a broader operational interpretation as the optimal reversible map in various contexts, such as approximate quantum error correction and quantum computation \cite{Barnum2000, Prabha10, Prabha24, Beigi16}, thermodynamic reversibility \cite{Alhambra17, Kwon19, Aw21, Scarani24}, and the study of quantum Markov chains and conditional independence \cite{Sutter16,Fawzi2015}. Despite its theoretical significance, the Petz map assumes knowledge of both the noisy channel and some encoded information about the input state to be recovered, often known as \textit{prior} \cite{Crooks2008, BuscemiScarani2021}. This dependence limits its practical applicability in scenarios where either the channel is only partially characterized or the initial state is unknown or inaccessible.

Previous works have investigated the role of non-ideal Petz maps and how this leads to suboptimal recovery, especially under strong or asymmetric noise~\cite{Lea22, Lea24}. Other approaches have also been proposed to improve the recovery performance of the Petz map. For instance, rotated Petz maps sharpen the data-processing inequality by averaging over a family of unitary rotations, introducing an additional layer of complexity~\cite{Sutter16, Renner}. Another attempt to enhance the recovery performance consists of averaging Petz maps~\cite{Barnum2000}, which aims to mitigate dependence on the unknown reference state. Conceptually, most existing refinements of the Petz recovery map remain single-shot: they attempt to optimize a single recovery step by modifying the reference state or introducing auxiliary optimization parameters. A recent exception is the work in~\cite{Murk2026}, which establishes a connection between successive Petz updates and maximum likelihood quantum tomography.  
A practical implementation of the Petz map has historically been challenging, given its (non-linear) dependence on the dual of the channel and the reference state. Moreover, it requires precise knowledge of the prior. However, recent progress has begun to bridge this gap. Explicit circuit constructions have been developed, allowing for an implementation with moderate gate complexity and auxiliary-system size overhead \cite{Wilde22}. On the experimental side, implementations have been realized on near-term platforms \cite{singh2025, Scarani25, Li2026}. Despite these advances, implementations remain tailored to specific experimental platforms, low-dimensional systems, and require accurate knowledge of the reference state. 

In this work, we introduce the \textit{Iterative Petz Recovery} (IPR) protocol, an adaptive strategy for recovering an unknown input state from multiple copies of its noisy output. Each copy passes through the noise channel once. Starting from the maximally mixed reference state, the protocol applies a sequence of Petz maps in which the output of each recovery step serves as the reference state for the next. In the finite-copy implementation, we estimate each updated reference from a fresh batch of noisy copies that have undergone the recovery steps constructed so far. The resulting sequence of maps is then applied to a retained noisy copy.

We characterize the recovery dynamics generated by the IPR and obtain exact analytical results for qubit depolarizing and single-axis Pauli channels. For these families of channels, we show that the iteration produces a systematic enhancement of the average recovery fidelity and characterize its asymptotic behavior. We further formulate a finite-copy implementation in which the intermediate reference states are reconstructed through quantum-state tomography and analyze the stability of the protocol under the resulting estimation errors. This provides an explicit connection between the ideal iterative dynamics and an implementation based on finite quantum resources.

Finally, we distinguish the convergence of the IPR from the more general possibility of learning and re-preparing an input state from multiple noisy copies. We compare finite-copy IPR with two tomography-based measure-and-prepare benchmarks, one using local Pauli tomography and one using collective tomography, and with the known optimal
deterministic estimator for depolarized pure qubits~\cite{Gendra2012}. The comparison does not establish a universal sample-complexity advantage for the IPR. Instead, it identifies its distinct operational structure: the estimated reference states determine a sequence of CPTP recovery maps acting on a retained quantum system, whereas measure-and-prepare inversion consumes the measured copies and produces a fresh, entanglement-breaking output. 

This work is organized as follows. In Sec.~\ref{sec:preliminaries} we establish the notation and the main mathematical tools used throughout the paper. In Sec.~\ref{sec:recovery} we discuss state recoverability and review the Petz recovery maps. Section~\ref{sec:ipr} presents the IPR protocol and our main results, including closed-form analytical  expressions for specific classes of quantum channels. For proofs and further details, we refer to the Appendix.

\section{NOTATION AND PRELIMINARIES}
\label{sec:preliminaries}

Throughout this work, we restrict ourselves to finite-dimensional Hilbert spaces, which we enumerate by capital letters. Quantum states on a quantum system $A$ are denoted by lowercase Greek letters, e.g., $\rho_A$. The set of all quantum states, i.e., all positive semidefinite trace-one matrices, is denoted by $\mathcal{D}$. Quantum channels are denoted by capital calligraphic letters, e.g., $\mathcal{N}^{A\leftarrow B}$. We omit system labels when they are clear from the context. The dimension of a quantum system $A$ is denoted by $|A|$. The main figure of merit to compare two quantum states $\rho$ and $\sigma$ will be the (squared) Uhlmann fidelity given by
\begin{equation}
    F(\rho,\sigma)=\norm{\sqrt{\rho}\sqrt{\sigma}}_1^2\,. 
\end{equation}
In particular, we will consider the Haar-averaged input-output fidelity of a quantum channel $\mE$ defined as
\begin{equation}
\label{eq:haar_avg}
\overline{F}(\mE):=\int d\mu(\psi)\,F\!\left(\psi,\mE(\psi)\right),
\end{equation}
where $d\mu(\psi)$ is the normalized Haar measure over pure states~\cite{Mele2024}. The Haar-averaged fidelity therefore quantifies the expected input-output fidelity for a uniformly distributed pure input and will serve as a benchmark for comparing the recovery channels and protocols considered below. Throughout this work, we will additionally make use of two standard norms, namely the trace norm defined as $\|X \|_1 := {\rm Tr}{\sqrt{X^\dagger X}}$ and the operator norm defined as $\|X\|_\infty := \sup_{\|\ket\psi\|=1}\|X\ket{\psi}\|$.

\section{State Recovery}
\label{sec:recovery}

Let $\mathcal{H}_A$ and $\mathcal{H}_B$ be finite-dimensional Hilbert spaces, and let $\mathcal{N}: \mathcal{L}(\mathcal{H}_A) \to \mathcal{L}(\mathcal{H}_B) $ be a quantum channel. Let $\rho, \sigma$ be quantum states and let $S(\rho\|\sigma) := \mathrm{Tr}(\rho\log\rho - \rho\log\sigma)$ denote the Umegaki quantum relative entropy \cite{Uhlmann77, Wilde15, Wilde2015} with ${\rm supp}(\rho) \subseteq {\rm supp}(\sigma)$, and $S(\rho\|\sigma)=\infty$ otherwise. The quantum relative entropy cannot increase under the application of a quantum channel $\mathcal{N}$, i.e., it satisfies the \textit{data processing inequality}
\begin{equation}
    \label{eq:relEnt}
    S(\rho\|\sigma) \geq S(\mathcal{N}(\rho)\|\mathcal{N}(\sigma)). 
\end{equation}
The equality holds if and only if the channel $\mathcal{N}$ is \textit{recoverable} with respect to the pair $(\rho, \sigma)$, meaning that there exists a CPTP recovery map $\mathcal{R}: \mathcal{L}(\mathcal{H}_B) \to \mathcal{L}(\mathcal{H}_A)$, such that $\mathcal{R}(\mathcal{N}(\rho)) = \rho$ and $\mathcal{R}(\mathcal{N}(\sigma)) = \sigma$~\cite{Petz1986, Petz1988}. For a fixed reference state $\sigma$, $\mathcal{R}$ is the Petz recovery map defined by 
\begin{equation}
    \label{eq:petz}
    \mathcal{P}^{\sigma}_{\mathcal{N}}(.) = \sigma^{\frac{1}{2}} \mathcal{N}^{\dagger}(\mathcal{N}(\sigma)^{-\frac{1}{2}} \,(.)\, \mathcal{N}(\sigma)^{-\frac{1}{2}})\sigma^{\frac{1}{2}}\,, 
\end{equation}
 $\mathcal{N}^\dagger$ denotes the adjoint channel with respect to the Hilbert-Schmidt inner product. This is a CPTP map (on the support of $\mathcal{N}(\sigma)$)~\cite{Petz1986,Petz1988}. The Petz recovery map can be interpreted as the quantum analogue of Bayes' rule, implementing an optimal reversal of the channel $\mathcal{N}$ conditioned on the prior state $\sigma$ \cite{Petz1986, Petz1988}. In this sense, $\sigma$ plays the role of a quantum prior, while $\mathcal{P}^{\sigma}_{\mathcal{N}}$ performs Bayesian inversion of the channel dynamics \cite{Crooks2008, BuscemiScarani2021}. Importantly, recovery should not be interpreted as a perfect inversion of the noise process itself, but rather as a reconstruction of a specific input state using the information encoded in the reference state $\sigma$. If $\sigma$ already encodes relevant information about the input state $\rho$, the corresponding Petz map typically achieves high recovery fidelity. The inverse appearing in Eq.~\eqref{eq:petz} becomes increasingly sensitive as the smallest nonzero eigenvalue of $\mathcal N(\sigma)$ approaches zero. Although a rank-deficient $\mathcal N(\sigma)$ can be treated using a generalized inverse on its support, our analysis requires ordinary inverses and explicit positive lower bounds on the smallest eigenvalues of both $\sigma$ and $\mathcal N(\sigma)$.
 
 Intuitively, the prior encodes our pre-existing bias about which inputs are likely. When the reference state poorly approximates the true input, the normalization factors $\mathcal{N}(\sigma)^{-\frac12}$ can strongly amplify noise. As a result, even small mismatches in $\sigma$ can lead to a stark degradation of recovery fidelity. The explicit dependence of the Petz map on its reference state raises an important practical question: how should this state be chosen when no prior information about the state to be recovered is available? A natural, unbiased choice is the maximally mixed state, which assigns no preferred direction in Hilbert space. Starting from this state, we introduce the iterative Petz recovery (IPR) protocol, in which the reference state is updated self-consistently at each iteration. The resulting procedure requires no state-specific choice of the initial reference and places the iterative protocol at the center of the recovery strategy.

\section{ITERATIVE PETZ RECOVERY}
\label{sec:ipr}

When no informative reference state is available, Petz recovery may initially appear inapplicable. Its explicit analytical form, however, allows this reference-state dependence to be incorporated into an adaptive strategy that progressively refines an unbiased/uninformative reference state, without requiring an optimization over recovery channels. Thus, every recovery step is explicitly defined and, under the relevant support conditions, constitutes a valid quantum channel.

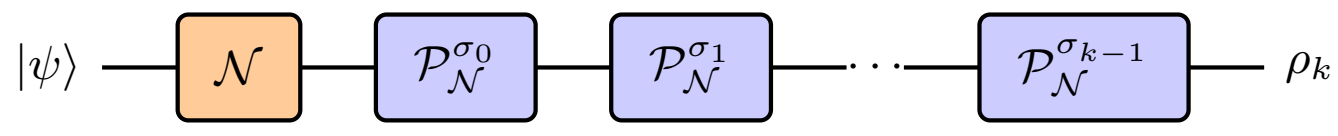
\begin{figure}[h!]
\centering
   \begin{quantikz}
   \lstick{$|\psi\rangle$} & \gate[wires=1,style=noise]{\,\,\mathcal{N}\,\,} & \gate[wires=1,style=recovery]{\,\,\mathcal{P}_\mathcal{N}^{\sigma_0}\,\,} & \gate[wires=1,style=recovery]{\,\,\mathcal{P}_\mathcal{N}^{\sigma_1}\,\,} & \push{\cdots} &\gate[wires=1,style=recovery]{\,\,\mathcal{P}_\mathcal{N}^{\sigma_{k-1}}\,\,} &\qw\rstick{$\rho_k$}
   \end{quantikz}
   \caption{Iterative Petz recovery (IPR) protocol that consists of sequential applications of the Petz map with updated reference states.}
   \label{fig:scheme}
\end{figure}

\begin{definition}
\label{def:ipr}
Let $\psi \in \mathcal{D}(\mathcal{H})$ be an unknown quantum pure state, and let $\mathcal{N}: \mathcal{L}(\mathcal{H}) \to \mathcal{L}(\mathcal{H})$ be a quantum channel, such that $\mathcal{N}(\psi)\neq \psi$. The \textit{Iterative Petz Recovery} protocol (IPR) generates two sequences of states: the state being processed at the $k$th step $\{\rho_k\}_{k\geq1}$ and the reference state used to define the corresponding Petz recovery map $\{\sigma_k\}_{k\geq1}$. Assume that both $\sigma_k$ and $\mathcal N(\sigma_k)$ are full rank, and set
\begin{align}
    \rho_0 &:= \mathcal{N}(\psi) \quad \text{and} \quad \sigma_0 = \mathbb{1}/d, 
\end{align}
for $d = \dim(\mathcal{H})$. For $k\in \mathbb{N}$, the IPR is defined recursively by 
\begin{align}
    \rho_{k + 1} := \mathcal{P}^{\sigma_{k}}_\mathcal{N}(\rho_k) \quad \text{and} \quad \sigma_{k + 1} := \rho_{k+1}.
\end{align}
\end{definition}
Notice that we exclude fixed points of the channel $\mathcal{N}(\psi) = \psi $. Here the noisy output already coincides with the initial state, and no recovery operation is required. We can extend the IPR to cover this case, by adopting the convention that the protocol terminates at initialization, and we assign recovery fidelity one at every iteration depth. The key feature of this protocol is its recursive structure: at each iteration, the reference state of the Petz map is updated to the state recovered in the previous step as depicted in Fig.~\ref{fig:scheme}. This adaptive update enables the protocol to refine its estimate of the original state, effectively updating the outputs at each step to improve the recovery fidelity in subsequent steps.

\subsection{Analytical average fidelity for depolarizing and
single-axis Pauli channel}\label{sec:analytics}

Having introduced the IPR protocol in general terms, we now investigate its average performance for two canonical classes of unital qubit noise: the depolarizing channel and single-axis Pauli channels. These models represent two distinct symmetry regimes: full rotational covariance and axial symmetry, respectively. Therefore, they provide a natural setting in which to isolate how the symmetry of the noise shapes the recovery performance. Importantly, these symmetries also allow us to obtain analytical results and to establish a direct relation between the Haar average and an average over a finite set of input states.

A pure qubit input state in the Bloch representation can be written as
\begin{equation}
\label{eq:pure_bloch}
    \psi(\vec r)=\frac{\mathbb{1}+\vec r\cdot\vec\sigma}{2},\quad \|\vec r\|_2=1,
\end{equation}
where $\|\cdot\|_2$ is the Euclidean norm on $\mathbb{R}^3$. We denote the recovery fidelity after the $k$th IPR step by 
\begin{equation}
\label{eq:ipr_directional_fidelity}
    F_k(\vec r):= {\rm Tr}\left[\psi(\vec r)\rho_k(\vec r) \right].
\end{equation}
Here, $\rho_k(\vec r)$ is the generally mixed recovered state corresponding to the pure input $\psi(\vec{r})$. To characterize the performance of the protocol across input directions, we consider the Haar average fidelity depicted in Eq.~\eqref{eq:haar_avg}. For the $k$th IPR step, the Haar average becomes
\begin{equation}
\label{eq:ipr_haar_average}
    \overline F_k:=\int d\mu(\psi)\,F_k(\vec r),
\end{equation}
For a fixed linear channel, this average can be evaluated exactly using a pure-state $2$-design~\cite{Bowdrey_02}. The average in Eq.~\eqref{eq:ipr_haar_average} is then replaced by 
\begin{equation}
\label{eq:ipr_pauli_design_average}
    F_k^{(2)}:=
    \frac{1}{6}\sum_{\vec r\in\mathcal S_{\mathrm P}}F_k(\vec r),\quad \mathcal S_{\mathrm P} :=
    \{\pm\vec e_x,\pm\vec e_y,\pm\vec e_z\}\,. 
\end{equation}
The average is taken over the six Pauli eigenstates. Such an equivalence, however, cannot be assumed a priori for the IPR, owing to the state-dependent nonlinearities introduced by the iterative update. Remarkably, the symmetries of the noise models considered here restore a precise connection between the two averages. Consider the depolarizing noise channel 
\begin{equation}
    \label{eq:depol}
    \mathcal{D}_{p}(\rho)=(1-p)\rho+p \,\Tr[\rho]\frac{\mathbb{1}}{2} ,\quad
    0\leq p < 1\,
\end{equation}
where $p$ denotes the noise strength. For the case of depolarizing noise, the average recovery fidelity can be analytically determined, which we do in the following Theorem.

\begin{theorem}\label{thm:depolarizingAverageFidelity}
Let $\mathcal{D}_p$ be the qubit depolarizing channel with $0\leq p < 1$. For every pure input state $\psi$, the fidelity generated by the IPR using the maximally mixed initial reference state is independent of the input $\psi$ and satisfies $\overline{F}_k=F_k^{(2)}=F_k(\psi)=F_k$, where
\begin{equation}\label{eq:depolarizingRecurrence}
F_{k+1}\!=\!F_k\left[\frac{(2\!-\!p)F_k}{2(1\!-\!p)F_k+p}\!+\!\frac{p(1\!-\!F_k)}{2-2(1\!-\!p)F_k-p}\right],
\end{equation}
with $F_1=1-p+\frac{p^2}{2}$. The sequence $\{F_k\}_{k\geq1}$ is strictly increasing and converges to unity. Moreover, its Haar average and $2$-design average coincide $\overline{F}_k=F_k^{(2)}=F_k(\psi)$. Furthermore,
\begin{equation}
\label{eq:depolarizing_convergence_rate}
\lim_{k\rightarrow\infty}\frac{1-F_{k+1}}{1-F_k}=\frac{p}{2-p},
\end{equation}
and hence the convergence is asymptotically geometric. For $p=0$, one has $F_k=1$ for all $k$.
\end{theorem}

This theorem provides a complete analytical characterization of the IPR for depolarizing noise. The recurrence~\eqref{eq:depolarizingRecurrence} therefore determines the full recovery performance at any finite $k$, without requiring an integration over input states. The theorem further shows that the IPR monotonically approaches perfect recovery for every $p<1$, including arbitrarily strong but non-complete depolarization. Eq.~\eqref{eq:depolarizing_convergence_rate} identifies that each additional iteration asymptotically removes a fixed fraction of the remaining infidelity. The convergence becomes progressively slower as $p\to1$, consistently with the approach to the completely depolarizing channel, for which all information about the input state is lost.

Now consider a single-axis Pauli noise, 
\begin{equation}
\label{eq:single_axis_channel}
    \mathcal{E}_{\alpha,p}(\rho)=(1-p)\rho+p\,P_\alpha\rho P_\alpha,
    \quad \alpha\in\{x,y,z\},
\end{equation}
where $0\leq p<1/2$ and $P_x = X, P_y = Y, P_z = Z$ are the Pauli operators. A paradigmatic example of this channel is dephasing in the computational basis, generated by stochastic $Z$ errors, which suppresses quantum coherences while leaving the populations unchanged. Analogously, stochastic $X$ and $Y$ errors describe bit-flip and bit-phase-flip noise along the corresponding axes $\alpha=x$ and $\alpha=y$. For a single-axis Pauli channel, axial symmetry reduces the Haar average to a one-dimensional integral. The Pauli $2$-design average is therefore exactly the Simpson approximation to the one-dimensional Haar integral~\cite{Atkinson1989}. Although this does not imply equality with the Haar average in general, it gives the finite average a precise symmetry-based interpretation and reduces its evaluation to three representative input sectors. The derivation and the corresponding quadrature error are discussed in Appendix~\ref{ap:analytics}. This symmetry reduction leads to the following exact characterization of the Pauli $2$-design fidelity. 

\begin{theorem}
\label{thm:singleAxisAverageFidelity}
Let $\mathcal{E}_{\alpha,p}$ be a single-axis Pauli channel with $0\leq p<1/2$. The Pauli $2$-design average fidelity generated by the IPR is
\begin{equation}
\label{eq:ap:avg2design_axis}
    F^{(2)}_k=\frac{1+2a_k}{3},
\end{equation}
where
\begin{equation}
\label{eq:singleAxisRecurrence}
a_{k+1}=a_k\left[\frac{(1-p)a_k}{a_k+p-2a_kp}+\frac{p(1-a_k)}{1-a_k-p+2a_kp}\right],
\end{equation}
with $a_1=1-2p(1-p)$. For $0<p<1/2$, the sequence $\{a_k\}_{k\geq1}$ is strictly increasing and converges to unity. Hence,
\begin{equation}
    \lim_{k\rightarrow\infty}F^{(2)}_k=1.
\end{equation}
Moreover,
\begin{equation}
\label{eq:singleAxisConvergenceRate}
    \lim_{k\rightarrow\infty}\frac{1-F^{(2)}_{k+1}}{1-F^{(2)}_k}=\frac{p}{1-p},
\end{equation}
corresponding to an asymptotically geometric convergence. For $p=0$, one has $F^{(2)}_k=1$ for all $k$.
\end{theorem}

\begin{figure}[ht]
    \centering
    \includegraphics[width=0.9\linewidth]{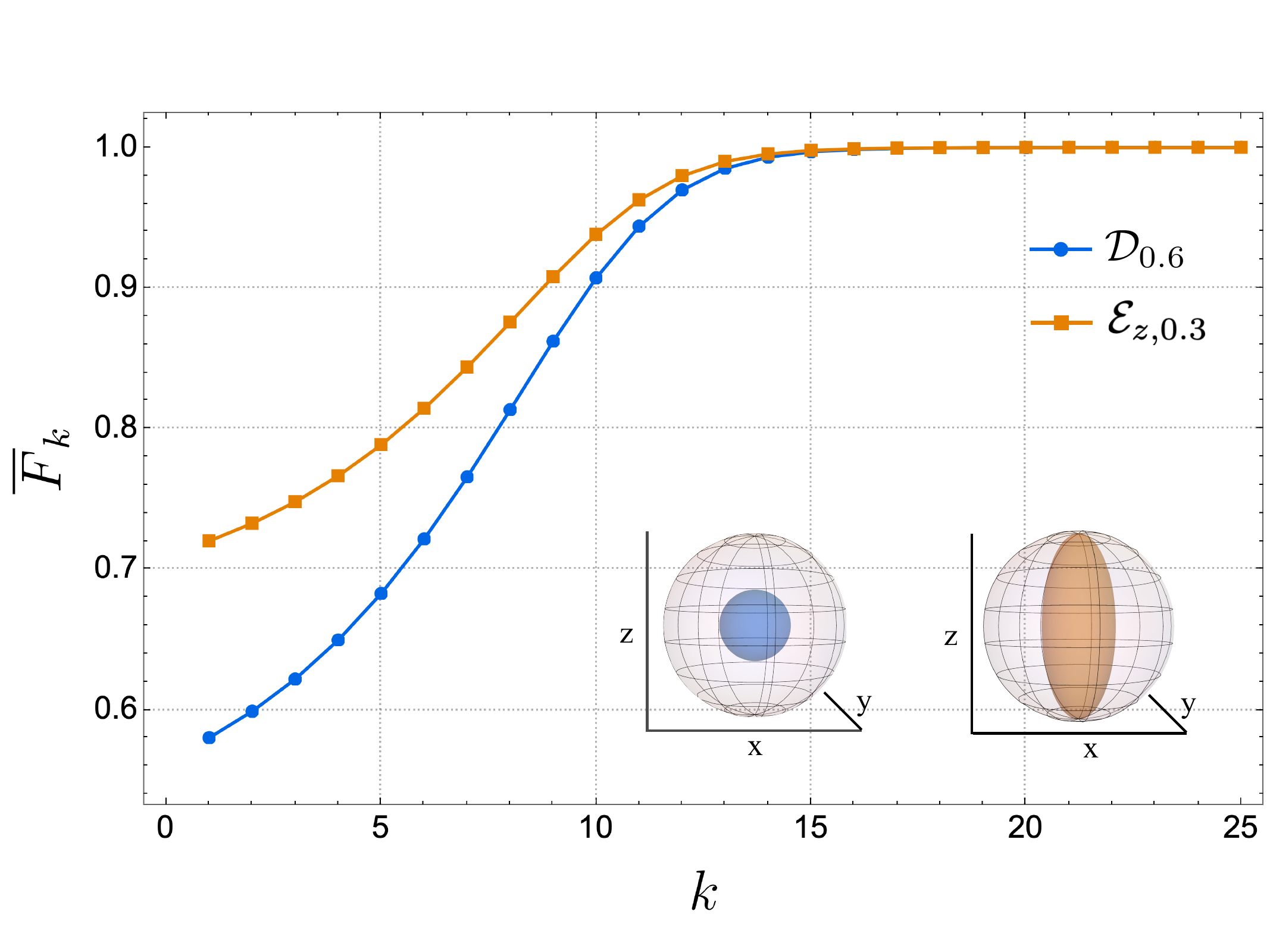}
    \caption{Average fidelity computed over pure qubit input states, for the depolarizing and dephasing channels as a function of the number of iterations of the IPR. Different noise structures lead to distinct convergence rates. The insets illustrate the corresponding actions on the Bloch sphere: depolarizing channel with $p = 0.6$ (blue), and dephasing channel with $p= 0.3$ (orange).}
    \label{fig:favg}
\end{figure}

Equation~\eqref{eq:ap:avg2design_axis} has a direct symmetry-based interpretation. Among the six Pauli eigenstates forming the $2$-design, the two states aligned with the noise axis are invariant under the channel and therefore retain unit fidelity throughout the protocol. The remaining four states lie along directions transverse to the noise axis and, by axial symmetry, share the same recovery fidelity $a_k$. The recurrence~\eqref{eq:singleAxisRecurrence} shows that, for $0<p<1/2$, every IPR step strictly improves the recovery of the transverse states. Since $a_k\to1$, the protocol asymptotically restores not only the states aligned with the noise axis, which are unaffected from the outset, but also the four initially degraded states in the design. It follows that the Pauli $2$-design average converges monotonically to unit fidelity. Similarly to the depolarizing case, recovery is rapid for weak noise but becomes progressively slower as $p\to1/2$. This behavior is physically consistent with the approach to complete dephasing at $p=1/2$, where the transverse Bloch components, and therefore the corresponding phase information, are entirely erased. The monotonic convergence to unity in Theorem~\ref{thm:singleAxisAverageFidelity} does not extend to the Haar average. The Simpson quadrature error depends on the iteration depth and need not vanish as $k\to\infty$. Numerical evaluation shows that the Haar average can become weakly nonmonotonic at large $k$ and approaches a value slightly below unity.

Figure~\ref{fig:favg} shows the recovery enhancement over the displayed finite range of iteration depths. For the
depolarizing channel, the Haar average coincides with the
$2$-design average and converges monotonically to unity. For
single-axis Pauli noise, the plotted Haar average initially
tracks the monotonically increasing $2$-design result very
closely. At larger iteration depths, however, the Haar average can exhibit a small decrease and converge to a value below unity, as discussed in Appendix~\ref{ap:analytics}.

\subsection{Finite-sample implementation of the IPR}

\begin{figure*}[t]
\centering
    \begin{quantikz}
    \lstick{$|\psi\rangle^{\otimes N_1}$} & \gate[wires=1,style=noise]{\mathcal{N}^{\otimes N_1}} & \gate[wires=1,style=recovery]{{\left(\mathcal{P}^{\sigma_0}_\mathcal{N}\right)}^{\otimes N_1}} & \meter{} \vcw{1} & & & & \\
    \lstick{$|\psi\rangle^{\otimes N_2}$} & \gate[wires=1,style=noise]{\mathcal{N}^{\otimes N_2}} & \gate[wires=1,style=recovery]{{\left(\mathcal{P}^{\sigma_0}_\mathcal{N}\right)}^{\otimes N_2}} & \gate[wires=1,style=recovery]{{\left(\mathcal{P}^{\hat{\sigma}'_1}_\mathcal{N}\right)}^{\otimes N_2}} & \meter{} & & & & \\
    \vdots &  &  &  &\vdots & & & \\
    \lstick{$|\psi\rangle$} & \gate[wires=1,style=noise]{\mathcal{N}} & \gate[wires=1,style=recovery]{{\mathcal{P}^{\sigma_0}_\mathcal{N}}}& \push{\cdots}  & \gate[wires=1,style=recovery]{{\mathcal{P}^{\hat{\sigma}'_{K-2}}_\mathcal{N}}} & \gate[wires=1,style=recovery]{\mathcal{P}^{\hat{\sigma}'_{K-1}}_\mathcal{N}}&\qw\rstick{$\rho_K'$}
    \end{quantikz}
    \caption{Tomography-assisted finite-copy implementation of the IPR protocol. At the $k$th intermediate stage, $N_k$ fresh copies of the input state are passed through the noisy channel $\mathcal{N}$ and through the recovery sequence constructed up to that stage. The resulting output state $\rho_k'$ is used as the reference state $\sigma'_k:=\rho_k'$ for the next iteration. State tomography on the $N_k$ outputs produces an estimate $\hat{\sigma}'_k$, from which the subsequent Petz map $\mathcal{P}_{\mathcal{N}}^{\hat{\sigma}'_k}$ is constructed. Since the initial unbiased reference state $\sigma_0$ is known exactly, the first implemented state coincides with its ideal counterpart, $\rho'_1=\rho_1$ and $\sigma'_1=\sigma_1$. For $k>1$, the implemented states depend on all preceding tomographic estimates, so finite-sample errors propagate throughout the adaptive recovery sequence. Once the estimates $\hat{\sigma}'_1,\ldots,\hat{\sigma}'_{K-1}$ have been obtained, the complete recovery sequence is applied to an additional unmeasured input, producing the final state $\rho_K'$.}
    \label{fig:scheme2}
\end{figure*}
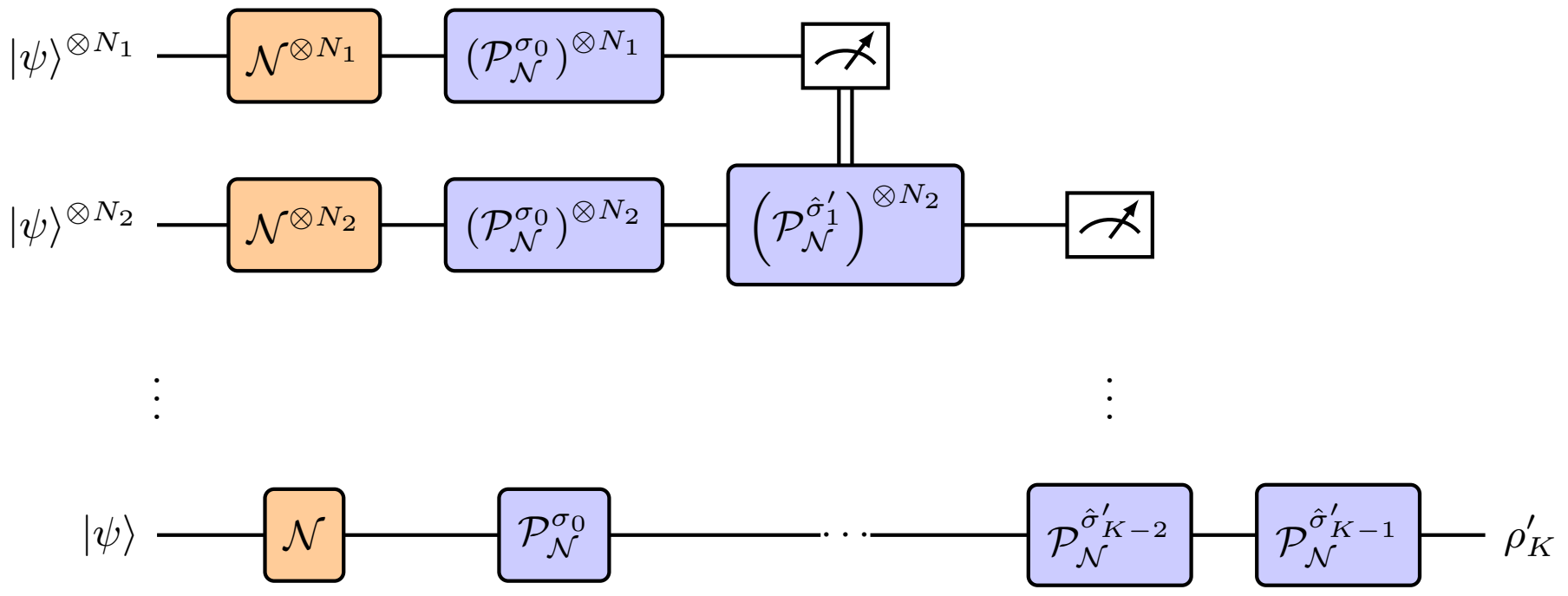

The IPR is formulated as a sequence of explicit recovery maps. However, its implementation requires more than a single realization of the maps. At the $k$th iteration, the update from $\rho_k$ to $\rho_{k+1}$ uses the Petz map constructed from $\sigma_k$. For $k \geq 1$, this reference state is the output of the preceding recovery step. Unless a classical description or an oracle preparing $\sigma_k$ is available, this state must be inferred from an ensemble of independently prepared realizations. The protocol must therefore be considered in a repeated-preparation setting, in which the preceding steps can be rerun to generate copies of the reference state required at each iteration. Each Petz map still acts on a single system. The additional copies constitute an informational resource used to construct that map. In an idealized implementation, one might assume sufficiently many copies to determine each reference state with arbitrary accuracy. A realistic implementation, however, must operate with a \textit{finite copy} budget.

Quantum algorithms for implementing the Petz recovery map have been developed under suitable access assumptions for the reference state and the noise channel, with their cost characterized in terms of circuit complexity and calls to these input resources~\cite{Wilde22, Prabha24}. Here, we address a complementary question: if access to the reference state is provided through \textit{state tomography}, how many copies are needed to construct the recovery map to a prescribed accuracy? More importantly, how do the resulting estimation errors propagate when the approximated constructed map is used to generate the reference state for the next iteration?

We distinguish the ideal IPR sequence $\{\rho_k,\sigma_k\}$ from the implemented sequence $\{\rho_k',\sigma'_k\}$. Here, $\rho_k'$ denotes the state produced at the $k$th implemented recovery step, while 
\begin{equation}
    \sigma'_k:=\rho_k'
\end{equation}
denotes the same state when regarded as the reference state for the subsequent Petz map. Its tomographic estimate is denoted by $\hat{\sigma}'_k$. For a target depth $K$, the protocol proceeds as follows.
\begin{enumerate}
    \item \textit{Initialization and copy allocation.} 
    Choose the known unbiased reference state $\sigma'_0=\sigma_0$ and allocate $N_k$ fresh copies of the input noisy state $\mathcal{N}(|\psi\rangle\langle\psi|)$ to estimate the reference state generated
    at each intermediate step $k=1,\ldots,K-1$.
    
    \item \textit{First recovery step.}
    Since $\sigma_0$ is known exactly, the first Petz map can be implemented without tomography. Applying it to the noisy input gives
    \begin{equation}
        \rho'_1 =
        \mathcal{P}_{\mathcal{N}}^{\sigma_0}
        \!\left(\mathcal{N}
        (|\psi\rangle\langle\psi|)\right), \quad
        \sigma'_1:=\rho'_1.
    \end{equation}
    Thus, in the absence of other implementation errors, the implemented
    and ideal sequences coincide at the first step.

    \item \textit{Tomographic estimation.}
    At stage $k$, generate $N_k$ copies of $\sigma'_k$ by passing $N_k$ fresh copies of the input state through the noise channel and through the recovery sequence constructed from $\sigma_0$ and the previous estimates
    $\hat{\sigma}'_1,\ldots,\hat{\sigma}'_{k-1}$.
    These outputs are consumed by state tomography, yielding the estimate
    $\hat{\sigma}'_k$. The corresponding estimation error is
    \begin{equation}
    \delta_k:=\left\|\hat{\sigma}'_k-\sigma'_k\right\|_\infty .
    \end{equation} 

    \item \textit{Construction of the estimated Petz map.}
    The estimator $\hat{\sigma}'_k$ is used to construct the Petz map
    \begin{equation}
        \mathcal{P}_{\mathcal{N}}^{\hat{\sigma}'_k}.
    \end{equation}

    \item \textit{Adaptive state update.}
    The next implemented state and reference state are defined recursively by
    \begin{align}
        \rho'_{k+1} &= \mathcal{P}_{\mathcal{N}}^{\hat{\sigma}'_k}\!\left(\rho_k'\right),
        \label{eq:implemented-rho-update}\\
        \sigma'_{k+1}&:=\rho'_{k+1}.
        \label{eq:implemented-sigma-update}
    \end{align}

    \item \textit{Iteration and final output.}
    Repeat steps 3--5 for $k=1,\ldots,K-1$. After the estimates $\hat{\sigma}'_1,\ldots,\hat{\sigma}'_{K-1}$ have been obtained, apply the resulting sequence of Petz maps to one additional, unmeasured noisy input.
\end{enumerate}

We denote the procedure above by $\mathcal R_{N, K}: \mathcal{L}(\mathcal{H}_B^{\otimes N}) \to \mathcal{L}(\mathcal{H}_A)$ the complete finite-copy implementation of the IPR under a total tomographic budget of $N$ state preparations. This notation encompasses the tomographic estimation of the successive reference states, the construction of the corresponding approximate Petz maps, and their sequential application. The final recovered state is 
\begin{equation}
    \rho'_{N,K} = \mathcal R_{N,K}\left[\mathcal{N}(\psi)^{\otimes N}\right]\,.
\end{equation}
When the number of iterations is fixed or clear from context, we write simply $\mathcal R_N$. Operationally, the copies used for tomography are consumed by the measurement. Therefore, each subsequent step is performed on a fresh batch of noisy inputs passed through the entire recovery sequence constructed up to that point. For a fixed tomographic budget $N$, the copies must be distributed among the intermediate stages
\begin{equation}
N=\sum_{k=1}^{K-1}N_k\,.
\end{equation}
See Fig.~\ref{fig:scheme2} for a schematic view of the implementation. 

The finite-copy implementation introduces statistical estimation errors at each step for which tomography is required. The deviations of $\{\rho_k', \sigma'_k\}$ from their exact-reference counterparts incorporate the errors accumulated over all preceding iterations. In what follows, we quantitatively compare the implemented protocol with the exact IPR sequence and determine how errors affect the recovery performance. 

\begin{theorem}
\label{th:ipr_fid}
Let $\psi$ be a pure state and let $\{\rho_k\}_{k\geq1}$ denote the output sequence generated by the exact-reference IPR. For a copy-budget $N$, let $\{\rho_k'\}_{k\geq1}$ denote the output sequence generated by the finite-copy implementation $\mathcal{R}_N$. Suppose that, at iteration $k$, 
\begin{equation}
    \|\rho_k'-\rho_k\|_1\leq \varepsilon_k.
\end{equation}
Then the fidelity achieved by the finite-copy implementation satisfies
\begin{equation}
\label{eq:fid_epsilon}
    F(\psi,\rho_k') \geq F(\psi,\rho_k)-\frac{1}{2} \varepsilon_k\,.
\end{equation}
\end{theorem}
Theorem~\ref{th:ipr_fid} gives the accumulated implementation error an immediate operational interpretation: at any iteration $k$, the reduction in recovery fidelity relative to the exact-reference IPR is at most $\varepsilon_k/2$. Hence, whenever the implemented state remains close to its exact-reference counterpart in trace distance, the recovery performance is preserved to the same accuracy. 

The error in Eq.~\eqref{eq:fid_epsilon} can be bounded in terms of the finite-copy resources used throughout the protocol. Appendix~\ref{ap:implement} combines the bounds derived in Appendix~\ref{ap:diamond_bounds}, which constitute a standalone result of independent interest valid for any single Petz map, with a tomographic guarantee to obtain a sufficient copy budget for the proposed construction, equivalently an upper bound on its minimum sample complexity.

\begin{proposition}
\label{prop:ipr_error_propagation}
Consider the exact-reference IPR sequences $\{\rho_k, \sigma_k\}_{k\geq1}$ and their finite-copy implementations $\{\rho_k', \sigma'_k\}_{k\geq1}$. Define the accumulated implementation error $\varepsilon_k:=\|\rho_k'-\rho_k\|_1$. Assume full-rank $d$-dimensional states satisfying 
\begin{equation}
    \lambda_{\min}(\sigma_k)\geq\mu_k,\quad \lambda_{\min}(\mathcal{N}(\sigma_k))\geq\nu_k,
\end{equation}
such that at every iteration under consideration,
\begin{equation}
    \delta_k+\frac{\varepsilon_k}{2}\leq\frac{1}{2}\min\left\{\mu_k,\frac{\nu_k}{\left\|\mathcal{N}(\mathbb{1})\right\|_{\infty}}\right\}.
\end{equation}
Here $\mu_k,\nu_k>0$ and $\lambda_{\min}$ denotes the minimal eigenvalue. Then, the accumulated implementation error obeys
\begin{equation}
    \varepsilon_{k+1}\leq\left(1+\frac{C_k}{2}\right)\varepsilon_k+C_k\delta_k\,,
\end{equation}
where 
\begin{equation}
    C_k := \frac{\sqrt{2}\,\|\mathcal{N}(\mathbb{1})\|_\infty}{\nu_k\sqrt{\mu_k}}+\frac{4\|\mathcal{N}(\mathbb{1})\|_\infty^2}{\nu_k^2}.
\end{equation}
\end{proposition}
Proposition~\ref{prop:ipr_error_propagation} provides an explicit finite-step error propagation bound for the IPR. The factor $1+C_k/2$ controls the transmission and possible amplification of errors inherited from previous iterations, while $C_k\delta_k$ quantifies the new error introduced by the tomographic estimation performed at iteration $k$. The dependence of $C_k$ on the lower spectral bounds $\mu_k$ and $\nu_k$ makes explicit the increasing sensitivity of the Petz map as either $\sigma_k$ or $\mathcal{N}(\sigma_k)$ approaches the boundary of the state space. The finite-sample bounds obtained here are understood on the regime for which $\mathcal{N}(\sigma_k)$ and $\mathcal N(\hat{\sigma}_k)$ remain full rank. Together with Theorem~\ref{th:ipr_fid}, this provides a direct quantitative connection between the available copy budget, the spectral conditioning of the successive Petz maps, and the final recovery fidelity.

\subsection{Comparison with measure-and-prepare recovery}

Unlike a fixed recovery channel acting on a single noisy input, the proposed finite-copy implementation of the IPR uses an ensemble of noisy states to estimate successive reference states and adapt the recovery map throughout the protocol. Any advantage over a single-copy recovery strategy must therefore be interpreted in light of these additional resources. Here, we take the number $N$ of available noisy copies as the primary resource and compare the IPR with $N$-copy measure-and-prepare (MP) recovery strategies. A complete resource comparison would additionally have to account for the number and complexity of the quantum operations required by each protocol, which we leave beyond the scope of the present analysis.

\begin{figure}[t]
    \centering
    \includegraphics[width=\linewidth]{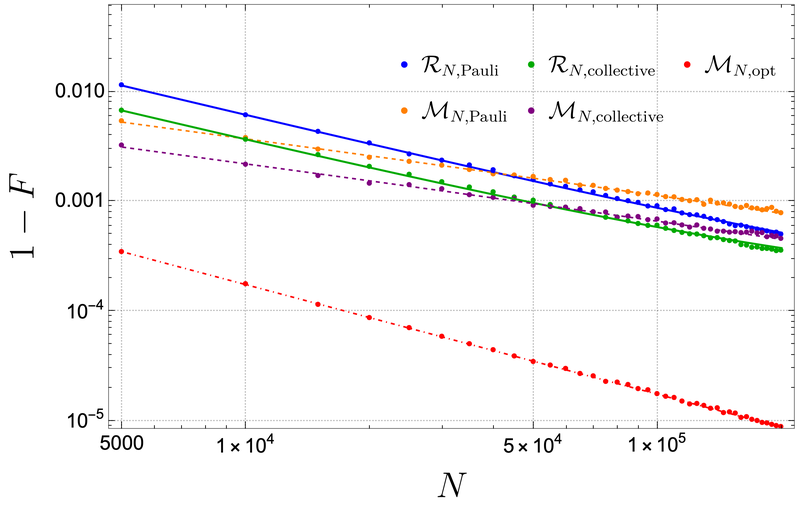}\\
    \includegraphics[width=\linewidth]{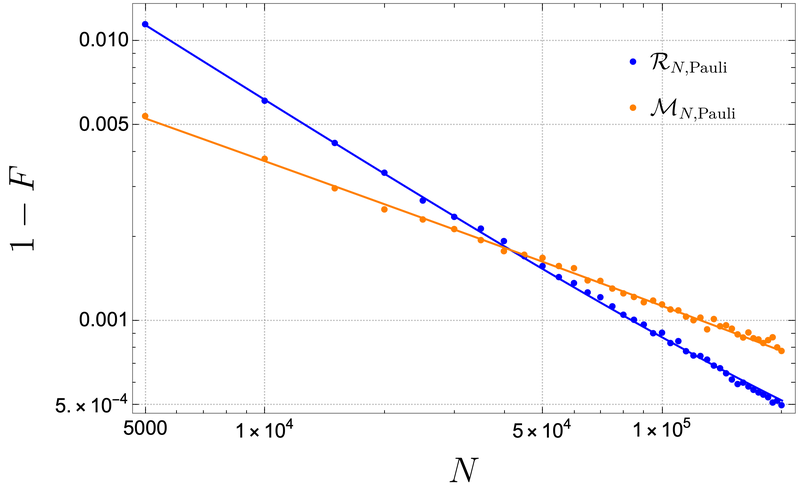}
    \caption{Finite-copy comparison of the implemented iterative Petz recovery protocol $\mathcal{R}_N$ with tomography-based and optimal MP benchmarks. The infidelity $1-F$ is plotted against the total tomographic copy budget $N$ on logarithmic axes for a depolarizing channel with $p=0.3$ and input state $|\psi\rangle=|+\rangle$. The IPR is evaluated after $K = 10$ IPR iterations, with the copies distributed uniformly among the $K-1$ tomographic stages. For the MP benchmarks, all $N$ independently prepared noisy copies are measured, the estimated Bloch vector is then classically inverted, projected radially onto the physical Bloch ball, and used to prepare a fresh output state. The fidelity is averaged over $2\times10^{3}$ independent tomographic trajectories. Upper panel: IPR and MP employing Pauli tomography (blue and orange, respectively) and collective tomography (green and purple, respectively). The optimal deterministic MP strategy is shown in red. Lower panel: IPR and MP employing Pauli tomography. The thick lines show empirical power-law fits over the displayed finite-copy range.}
    \label{fig:strategy_multi}
\end{figure}

An $N$-copy MP protocol measures the noisy ensemble, and prepares an output state conditioned on the measurement outcome. Its general form is 
\begin{equation}
\label{eq:MIP_channel}
\mathcal{M}_N(\rho^{\otimes N}) =\sum_m {\rm Tr}\left[M_m\rho^{\otimes N}\right]\tau_m ,
\end{equation}
where $\rho = \mathcal{N}(\psi)$ is the noisy state, $\{M_m\}_m$ is a POVM on the $N$-copy input space, and $\tau_m$ is the state prepared upon obtaining outcome $m$. Depending on the available experimental capabilities, the measurement may be implemented independently on the individual copies or collectively on the complete ensemble~\cite{Cirac1999, Bagan2006}. 

Within this class, we consider the following specific MP strategy: use the measurement outcomes to reconstruct an unrestricted mixed-state estimate of the noisy input, apply the classical inverse of the known noise transformation to the estimated state parameters, project the result onto the physical state space when necessary, and prepare the resulting state. We consider two benchmarks: $\mathcal{M}_{N,\rm Pauli}$, based on independent Pauli measurements, and $\mathcal{M}_{N, \rm collective}$, based on a collective measurement of the noisy ensemble. In both cases, the prepared states $\tau_m$ in Eq.~\eqref{eq:MIP_channel} are determined by the corresponding estimation, inversion, and projection procedure. 

To place these two baselines in the context of the full MP class, we additionally include $\mathcal{M}_{N,\rm opt}$, the optimal deterministic MP strategy for estimating the original pure state from $N$ depolarized copies~\cite{Gendra2012}. Unlike the two tomography-based baselines, this strategy does not first reconstruct an arbitrary mixed state and subsequently invert the channel. Instead, it uses prior knowledge that the original state is pure, knowledge of the depolarizing parameter $p$, a collective measurement, and a guessing rule that directly maximizes the final average fidelity. Consequently, $\mathcal{M}_{N,\rm opt}$ is an ideal, task-specific benchmark, whereas $\mathcal{M}_{N,\rm Pauli}$ and $\mathcal{M}_{N,\rm collective}$ represent generic MP procedures. The rotational covariance of the depolarizing channel makes the performance of $\mathcal{M}_{N,\rm opt}$ independent of the input direction, allowing a direct comparison with the results for $\ket{\psi}=\ket{+}$.

Fig.~\ref{fig:strategy_multi} compares the infidelity achieved by the finite-copy IPR and MP benchmarks as a function of the total tomographic copy-budget for the depolarizing channel. For the IPR the copies are distributed approximately uniformly among the tomographic stages $N_k\simeq N/(K-1)$. The sensitivity to alternative allocations is illustrated in Appendix~\ref{ap:comparison}. The upper panel shows the comparison for both Pauli and collective tomographic strategies. The thick lines represent empirical power-law fits over this finite-copy interval and should not be interpreted as asymptotic scaling laws. The two tomography-based MP strategies first estimate the noisy input as an unrestricted mixed state before applying the inverse channel and projecting the reconstructed state onto the Bloch ball. After inversion, statistical fluctuations of order $N^{-1/2}$ in the estimated Bloch-vector radius can move the reconstructed state across the pure-state boundary. This boundary effect produces an $N^{-1/2}$ contribution to the mean infidelity and accounts for the slower behavior observed for both $\mathcal{M}_{N, \rm Pauli}$ and $\mathcal{M}_{N,\rm collective}$. Collective measurements improve the statistical efficiency by exploiting the permutation and rotational symmetries of the full $N$-copy state. Their advantage here is therefore statistical rather than experimental: although the Schur transform underlying such measurements admits an efficient quantum-circuit implementation in principle~\cite{Bacon2006Schur}, it requires coherent operations across multiple copies and is substantially more demanding to realize than independent Pauli measurements~\cite{Hou2018Collective}. For the IPR, the fitted exponent of approximately $0.9$ is consistent with an approximately $N^{-1}$ finite-sample correction over the range considered and eventually outperforms the corresponding tomographic-inversion estimator. The optimal deterministic MP strategy shows the best performance attainable within the task-specific deterministic MP setting considered here.

The lower panel in Fig.~\ref{fig:strategy_multi} shows the comparison when both the IPR and MP employ independent Pauli tomography. At small copy budgets, $\mathcal{M}_{N,\rm Pauli}$ achieves the lower infidelity. As $N$ increases, however, the curves cross and the IPR attains the lower infidelity over the remaining displayed range. This faster empirical decay accounts for the observed crossover with the tomography-based MP strategies.  In particular, these crossovers demonstrate an advantage over the specified tomography-based inversion procedures, not over the optimal deterministic MP strategy or over MP protocols in general.

\section{Discussion and Outlook}
\label{sec:conclusions}
In this work, we introduced the iterative Petz recovery protocol as an adaptive approach to quantum state recovery. The construction turns the dependence of the Petz map on a reference state into an iterative mechanism: starting from a prescribed initial reference, each step generates an updated reference state that determines the recovery map used at the next iteration. The IPR thereby replaces a sequence of recovery optimizations with a sequence of explicit Petz maps whose reference states are progressively refined. We analyzed both the exact-reference dynamics of this protocol and its implementation when only finitely many copies of the noisy input are available. For qubit depolarizing and single-axis Pauli channels, we derived closed recurrence relations for the recovery fidelity averaged over pure input states. These results identify regimes in which the average fidelity improves monotonically with the iteration depth and show how the symmetries of the noise channel constrain the otherwise nonlinear iterative dynamics. In particular, they provide analytically tractable examples in which adaptive reference-state refinement leads to a systematic improvement over a single application of the Petz map. 

We then addressed the central implementation question of how the protocol performs when its intermediate reference states are known only through finite-sample tomography. At each iteration, the accuracy of the estimated reference state is determined by the number of copies assigned to that tomographic stage. The diamond-norm perturbation bounds derived in the Appendix translate these estimation errors into errors in the corresponding Petz maps under explicit spectral regularity conditions. Composing these bounds across iterations yields a finite-step estimate for the accumulated implementation error and, thus, a direct guarantee on the achieved recovery fidelity. This analysis identifies the tomographic errors and the spectral conditioning of the intermediate reference states as the main factors governing the robustness of the finite-copy IPR. The Appendix additionally quantifies the sensitivity of the Petz map to imperfect knowledge of the noise channel.

To place the finite-copy IPR in the broader context of multi-copy recovery, we compared it with several measure-and-prepare (MP) strategies. Over the copy range investigated for the depolarizing channel, the IPR exhibits a steeper empirical reduction of the infidelity than the tomography-based MP procedures. This behavior allows the IPR to outperform these generic tomographic-inversion strategies beyond the observed finite-copy crossover. We therefore do not interpret our comparison as establishing a universal sample-complexity advantage for the IPR, nor do we infer asymptotic scaling from the finite-range numerical fits. Instead, the comparison highlights a structural distinction between the two approaches. An MP protocol measures and consumes the available copies before preparing a new output state and is therefore entanglement breaking with respect to its measured input. By contrast, the IPR uses part of the ensemble to determine a sequence of explicit CPTP recovery maps, which are then applied directly to a retained noisy system. Our results establish iterative reference-state refinement as an analytically tractable recovery principle and provide a first quantitative account of its finite-resource requirements.

There are several directions for future research. A first objective is to determine whether the observed monotonicity and convergence properties extend beyond the Pauli channels studied here and to obtain a general characterization of the fixed points and convergence rates of the iterative dynamics. On the implementation side, it will be important to optimize the allocation of a fixed copy budget among the tomographic stages and to incorporate circuit complexity, gate imperfections, and uncertainty in the noise model into a unified resource analysis. The behavior near rank-deficient reference states, where the Petz map becomes increasingly sensitive to estimation errors, also calls for perturbed implementations and sharper perturbation bounds. Further comparisons should include channel-adapted and minimax MP benchmarks for anisotropic noise, as well as recovery tasks in which preserving correlations with an external reference is part of the performance criterion. Finally, extending the IPR to bosonic Gaussian systems would require determining whether the iterative dynamics remain closed within the Gaussian setting, such that the successive reference states and recovery maps can be characterized entirely through first moments and covariance matrices. Such a formulation would provide a tractable framework for studying adaptive recovery under experimentally relevant loss and thermal-noise channels.

\section*{Acknowledgments}

LL thanks N. K. Bernardes and M. T. Quintino for useful discussions. FA thanks his cat Lina for moral support. MBP thanks cat Ayala for lllkjuztfdas. During the preparation of this manuscript, the authors used ChatGPT (OpenAI; GPT-5.6 Sol) to assist with grammatical refinement and to cross-check selected intermediate calculations. All mathematical derivations, results, and AI-assisted revisions were independently reviewed and verified by the authors. Numerical calculations were performed with Wolfram Mathematica and MATLAB using CVX~\cite{cvx}. Quantum circuit diagrams were typeset using the Quantikz package~\cite{KayQuantikz}.

\appendix

\setcounter{lemma}{0}
\setcounter{theorem}{0}
\setcounter{corollary}{0}
\setcounter{proposition}{0}

\numberwithin{proposition}{section}

\section{Bounds on diamond distances between Petz maps}\label{ap:diamond_bounds}
In this section, we provide bounds on the so-called \textit{diamond distance} between two Petz maps, which we will use to bound the recovery fidelity. For quantum channels $\mN^{A\to B}$ and $\mM^{A\to B}$, the so-called diamond distance $\norm{\cdot}_\diamond$~\cite{Kitaev1997,Aharonov1998,Kitaev2002,Watrous2009,Watrous2013,Watrous2018} is defined as
\begin{align}\label{eq:def_DiamondDistance}
   \mkern-7mu \frac{1}{2}\norm{\mN\!-\!\mM}_{\diamond} \!:=\! \frac{1}{2}\sup_{\rho_{AR}}\! \norm{\idChan_R \otimes \left(\mN\!-\! \mM\right)(\rho_{AR})}_1 \,,
\end{align}
where the supremum over $\rho_{AR}$ is to be understood as an optimization over bipartite states with an auxiliary system $R$ of arbitrary dimension. For finite-dimensional systems, it suffices to consider auxiliary systems $R\cong A$, see~\cite{Belavkin2005, Puzzuoli2017}. 

To provide bounds on the diamond distance between two Petz channels with respect to different reference states and input channels, we need the following preliminary lemmas. We start with an upper bound on the diamond norm between two quantum operations with a single Kraus operator, which appeared in~\cite{Kretschmann2008,Watrous2009} before.
\begin{lemma}
\label{lem:DiamondBoundConjugation}
    Let $\mathcal{N}(X) := AXA$ and $\mathcal{M}(X) := BXB$, with $A,B \in \Herm$. Then, 
    \begin{align}
    \|\mathcal{N}-\mathcal{M}\|_\diamond \leq \left( \norm{A}_\infty+\norm{B}_\infty\right) \|A-B\|_\infty.
\end{align}
\end{lemma}
\begin{proof}
Recall that
\begin{align}
\|\mathcal{N}-\mathcal{M}\|_\diamond&:= \sup_{\rho_{AR}}\|\mathrm{id}_R\otimes (\mathcal{N}-\mathcal{M})(\rho_{AR})\|_1.
\end{align}
Let $\rho_{AR}$ denote an arbitrary state, where we omit the system labels from now on. Using the standard inequality~\cite{Watrous2018} 
\begin{align}\label{eq:trace_dist_ineq}
    \| XYZ\|_1 \le \|X\|_\infty\,\|Y\|_1\,\|Z\|_\infty,
\end{align}
together with $AXA-BXB=(A-B)XB+AX(A-B)$ yields
\begin{align}
&\|\mathrm{id}_R\otimes (\mathcal{N} \mkern-4mu- \mkern-4mu\mathcal{M})\!(\rho)\|_1 \nonumber \\
&\, \leq \|(A\mkern-4mu-\mkern-4muB)\!\otimes \!I_R\|_\infty\,\|\rho\|_1\,\|A^\dagger\otimes I_R\|_\infty \nonumber \\
&\quad +\|B\otimes I_R\|_\infty\,\|\rho\|_1\,\|(A-B)^\dagger\otimes I_R\|_\infty \nonumber \\
&=\bigl(\|A\|_\infty+\|B\|_\infty\bigr)\,\|A-B\|_\infty,
\end{align}
where we used that $\|X\otimes I_R\|_\infty= \|X\|_\infty$, and $\|\rho\|_1=1$. This holds for all feasible $\rho$, and thus, taking the supremum over $\rho$ on both sides completes the proof.
\end{proof}

\begin{lemma}\label{lem:infty_conti}
Let $A,B> 0$. Then,
\begin{subequations}
\begin{align}
    &\norm{A^{\tfrac{1}{2}}-B^{\tfrac{1}{2}}}_\infty\;\le\;\frac{1}{2\sqrt{\gamma}}\,\norm{A-B}_\infty, \\
    &\norm{A^{-\tfrac{1}{2}}-B^{-\tfrac{1}{2}}}_\infty\;\le\;\frac{1}{2\sqrt{\gamma^3}}\,\norm{A-B}_\infty,
\end{align}
\end{subequations}
where $\gamma :=\min \{\lambda_{\min}(A),\lambda_{\min}(B)\}$.
\end{lemma}
\begin{proof}
Consider the integral representation of the (inverse) square root of a matrix $A$ \cite[Chapter 6]{Horn1991}
\begin{subequations}
\begin{align}
    A^{\tfrac{1}{2}}&= \frac{1}{\pi} \int_{0}^{\infty} dt\, t^{-\tfrac{1}{2}} A(A+t\mathbb{1})^{-1}  \\
    A^{-\tfrac{1}{2}}&= \frac{1}{\pi} \int_{0}^{\infty} dt \,t^{-\tfrac{1}{2}} (A+t\mathbb{1})^{-1}.
\end{align}
\end{subequations}
Together with the identities
\begin{align}
    &(A\!+\!t\mathbb{1})^{-1} \!-\!(B\!+\!t\mathbb{1})^{-1}= (A\!+\!t\mathbb{1})^{-1} (B\!-\!A)(B\!+\!t\mathbb{1})^{-1},  \nonumber  \\
    &A(A\!+\!t\mathbb{1})^{-1}= \mathbb{1}\!-\! t (A\!+\!t\mathbb{1})^{-1} \nonumber,
\end{align}
we obtain 
\begin{align}
    &\norm{A^{\tfrac{1}{2}}-B^{\tfrac{1}{2}}}_\infty \nonumber \\
    &\leq \frac{1}{\pi} \int_{0}^{\infty} dt \, t^{\tfrac{1}{2}}  \norm{(B\!+\!t\mathbb{1})^{-1}}_\infty \norm{A\!-\!B}_\infty \norm{(A\!+\!t\mathbb{1})^{-1}}_\infty \nonumber \\
    &\leq \frac{1}{\pi} \int_{0}^{\infty} dt \, \frac{t^{\tfrac{1}{2}}}{(\gamma\!+\!t)^2}  \norm{A\!-\!B}_\infty  =\frac{1}{2\sqrt{\gamma}} \norm{A\!-\!B}_\infty,
\end{align}
and analogously
\begin{align}
    &\norm{A^{-\tfrac{1}{2}}-B^{-\tfrac{1}{2}}}_\infty \nonumber \\ 
    &\leq \frac{1}{\pi} \int_{0}^{\infty} \!dt \,t^{-\tfrac{1}{2}}  \norm{(B\!+\!t\mathbb{1})^{-1}}_\infty \norm{A\!-\!B}_\infty \norm{(A\!+\!t\mathbb{1})^{-1}}_\infty \nonumber\\
    &\leq \frac{1}{\pi} \!\int_{0}^{\infty}\!\! dt \, \frac{t^{-\tfrac{1}{2}}}{(\gamma\!+\!t)^2}  \norm{A\!-\!B}_\infty  \!=\!\frac{1}{2\sqrt{\gamma^3}} \norm{A\!-\!B}_\infty.
\end{align}
\end{proof}

Next, we begin to derive bounds on the recovery fidelity in terms of the diamond distance between Petz maps.

\begin{lemma}
Let $\psi$ be a pure state and let $\mathcal{N}$ be a quantum channel. The difference between the (recovery) fidelities between two Petz maps is bounded by
\begin{align*}
&\left|
F\!\left(\psi,\!\mathcal{P}_{\mathcal{N}}^\sigma\!\circ\!\mathcal{N}(\psi)\right)
\!-\!
F\!\left(\psi,\!\mathcal{P}_{\hat{\mathcal{N}}}^{\hat{\sigma}}\!\circ\!\mathcal{N}(\psi)\right)
\right|\leq
\frac{1}{2}\left\|\mathcal{P}_{\mathcal{N}}^\sigma-\mathcal{P}_{\hat{\mathcal{N}}}^{\hat{\sigma}}\right\|_{\diamond}.
\end{align*}
\end{lemma}
\begin{proof}
Since $\psi$ is pure, the fidelity reduces to the overlap $F(\psi,\rho)={\rm Tr}[\psi\rho].$ Now let
\begin{equation}
    \rho :=
    \mathcal{P}_{\mathcal{N}}^\sigma\circ\mathcal{N}(\psi),
    \qquad
    \tau :=\mathcal{P}_{\hat{\mathcal{N}}}^{\hat{\sigma}}\circ\mathcal{N}(\psi),
\end{equation}
then $\rho$ and $\tau$ are quantum states, and therefore, $\rho-\tau$ is Hermitian and traceless. Hence,
\begin{align}
\left|F(\psi,\rho)\!-\!F(\psi,\tau)\right|=\left|{\rm Tr}[\psi(\rho\!-\!\tau)]\right| \leq\tfrac{1}{2}\|\rho\!-\!\tau\|_1,
\end{align}
where the inequality follows from the variational characterization of the trace norm, see~\cite{Watrous2018}. Using the definitions of $\rho$ and $\tau$, yields
\begin{align}
\left|F(\psi,\rho)-F(\psi,\tau)\right|
&\leq
\frac{1}{2}
\left\|
\left(\mathcal{P}_{\mathcal{N}}^\sigma-\mathcal{P}_{\hat{\mathcal{N}}}^{\hat{\sigma}}\right)
\circ\mathcal{N}(\psi)
\right\|_1 \nonumber \\
&\leq
\frac{1}{2}
\left\|
\left(\mathcal{P}_{\mathcal{N}}^\sigma-\mathcal{P}_{\hat{\mathcal{N}}}^{\hat{\sigma}}\right)
\circ\mathcal{N}
\right\|_{\diamond}.
\end{align}
Finally, by the submultiplicativity of the diamond norm~\cite[Chapter 3]{Watrous2018}
\begin{equation}
\left\|
\left(\mathcal{P}_{\mathcal{N}}^\sigma-\mathcal{P}_{\hat{\mathcal{N}}}^{\hat{\sigma}}\right)
\circ\mathcal{N}
\right\|_{\diamond}
\leq
\left\|
\mathcal{P}_{\mathcal{N}}^\sigma-\mathcal{P}_{\hat{\mathcal{N}}}^{\hat{\sigma}}
\right\|_{\diamond},
\end{equation}
where we used that $\|\mathcal{N}\|_{\diamond}=1$ for the channel $\mN$.

\end{proof}

\begin{proposition}
\label{prop:main_bound}
Let $\mathcal{N},\hat{\mathcal{N}}:\mathcal{L}(\mathcal{H})\rightarrow \mathcal{L}(\mathcal{H})$ denote two channels, and let $\sigma,\hat{\sigma}$ be full-rank quantum states. Suppose $\mN(\sigma)$ and $\hat{\mN}(\hat{\sigma})$ are full-rank. Let $\mathcal{P}_\mathcal{N}^\sigma$ and $\mathcal{P}_{\hat{\mathcal{N}}}^{\hat{\sigma}}$ denote the corresponding Petz recovery maps. Then,
\begin{align}\label{eq:Dist_Petz_Bound}
     \|\mathcal{P}^\sigma_\mathcal{N} \!-\! \mathcal{P}^{\hat{\sigma}}_{\hat{\mathcal{N}}}\|_\diamond &\!\leq\!  \|\Sigma\|_\infty (\|\sigma^{\frac12}\|_\infty \!+\! \|\hat\sigma^{\frac12}\|_\infty)\,\|\mathcal{N}^\dagger\circ S_{\mathcal{N}(\sigma)^{-\frac12}}\|_\diamond\, \nonumber \\
    &+ \|\hat{\sigma}\|_\infty \|\hat{\mathcal{N}}^\dagger \circ (S_{\mathcal{N}(\sigma)^{-\frac12}} - S_{\hat{\mathcal{N}}(\hat\sigma)^{-\frac12}})\|_\diamond \nonumber\\
    &+ \|\hat{\sigma}\|_\infty  \|(\mathcal{N}^\dagger - \hat{\mathcal{N}}^\dagger) \circ S_{\mathcal{N}(\sigma)^{-\frac12}}\|_\diamond \,,
\end{align}
with $\Sigma = \sigma^{\frac12}- \hat{\sigma}^{\frac12}$ and $S_{\mathcal{N}(\sigma)^{-\frac12}}(\cdot) = \mathcal{N}(\sigma)^{-\frac12}(\cdot)\mathcal{N}(\sigma)^{-\frac12}$.
\end{proposition}
\begin{proof}
Recall that the Petz recovery maps can be written as
\begin{equation}
    \mathcal{P}_{\mathcal{N}}^{\sigma}=S_{\sigma^{1/2}}
    \circ\mathcal{N}^{\dagger}\circ S_{\mathcal{N}(\sigma)^{-1/2}},
\end{equation}
where the ``sandwich" maps are defined as $S_X(\rho)=X\rho X$, and analogously
\begin{equation}
    \mathcal{P}_{\hat{\mathcal{N}}}^{\hat{\sigma}}=
    S_{\hat{\sigma}^{1/2}}\circ\hat{\mathcal{N}}^{\dagger}\circ
    S_{\hat{\mathcal{N}}(\hat{\sigma})^{-1/2}}.
\end{equation}
The inverses are well defined since by assumption $\mN(\sigma)>0$ and $\hat{\mN}(\hat{\sigma})>0$. For convenience, let us define
\begin{subequations}
\begin{align}
    \Phi&:=\mathcal{N}^{\dagger}\circ S_{\mathcal{N}(\sigma)^{-1/2}}, \\
    \hat{\Phi} &:=\hat{\mathcal{N}}^{\dagger}\circ S_{\hat{\mathcal{N}}(\hat{\sigma})^{-1/2}}.
\end{align}
\end{subequations}
The two Petz maps therefore satisfy
\begin{subequations}
\begin{align}
    \mathcal{P}_{\mathcal{N}}^{\sigma}(X)&=\sigma^{\tfrac{1}{2}}\Phi(X)\sigma^{\tfrac{1}{2}}, \\
    \mathcal{P}_{\hat{\mathcal{N}}}^{\hat{\sigma}}(X)&=\hat{\sigma}^{\tfrac{1}{2}}\hat{\Phi}(X)\hat{\sigma}^{\tfrac{1}{2}}.
\end{align}
\end{subequations}
Adding and subtracting suitable intermediate terms lets us express the action the difference of Petz maps onto an arbitrary operator $X$ as
\begin{align}
\label{eq:petz_difference_split}
    \left(\mathcal{P}_{\mathcal{N}}^{\sigma}-\mathcal{P}_{\hat{\mathcal{N}}}^{\hat{\sigma}}\right)(X)=&
    \left(\sigma^{\tfrac{1}{2}}-\hat{\sigma}^{\tfrac{1}{2}}\right)\Phi(X)\sigma^{\tfrac{1}{2}} \nonumber \\
    &+\hat{\sigma}^{\tfrac{1}{2}}\Phi(X)\left(\sigma^{\tfrac{1}{2}}-\hat{\sigma}^{\tfrac{1}{2}}\right) \nonumber \\
    &+\hat{\sigma}^{\tfrac{1}{2}}\left(\Phi(X)-\hat{\Phi}(X)\right)\hat{\sigma}^{\tfrac{1}{2}}.
\end{align}
Using the definition of $\Sigma$, this becomes
\begin{align}
    \left(\mathcal{P}_{\mathcal{N}}^{\sigma}-\mathcal{P}_{\hat{\mathcal{N}}}^{\hat{\sigma}}\right)(X) &=\Sigma\Phi(X)\sigma^{\tfrac{1}{2}}+\hat{\sigma}^{\tfrac{1}{2}}\Phi(X)\Sigma\nonumber \\
    &+\hat{\sigma}^{\tfrac{1}{2}}\left(\Phi(X)-\hat{\Phi}(X)\right)\hat{\sigma}^{\tfrac{1}{2}}.
    \label{eq:petz_difference_sigma}
\end{align}
For arbitrary operators $A$ and $B$, let $\Xi_{A,B}$ denote the linear map acting as $\Xi_{A,B}(X)=AXB$. Then, for any linear map $\Lambda$, the diamond norm satisfies 
\begin{equation}
\label{eq:left_right_bound}
    \left\|\Xi_{A,B} \circ \Lambda \right\|_{\diamond} \leq\|A\|_{\infty}\|\Lambda\|_{\diamond}\|B\|_{\infty}.
\end{equation}
This follows directly from Eq.~\eqref{eq:trace_dist_ineq} by using that
\begin{align}
&\left\| (\Xi_{A,B}\! \circ \!\Lambda) \!\otimes\! \idChan_R (Z_{AR})\right\|_{1}\nonumber \\ 
&= \! \bigl\|(A\otimes I_R)(\Lambda\otimes\operatorname{id}_R)(Z_{AR})
                    (B\otimes I_R)\bigr\|_1 \nonumber\\
& \le
\|A\!\otimes\! I_R\|_\infty
\|\Lambda\!\otimes\!\operatorname{id}_R(Z_{AR})\|_1
\|B\!\otimes \!I_R\|_\infty \nonumber \\
&\leq
\|A\|_\infty \|\Lambda\|_\diamond\|B\|_\infty \norm{Z_{AR}}_1.
\end{align}
Taking the supremum over all feasible $Z_{AR}$ proves Eq.~\eqref{eq:left_right_bound}. Using this together with the triangle inequality (for the diamond norm) in Eq.~\eqref{eq:petz_difference_sigma} gives
\begin{align}\label{eq:diamond_triangle}
    \left\|\mathcal{P}_{\mathcal{N}}^{\sigma}-\mathcal{P}_{\hat{\mathcal{N}}}^{\hat{\sigma}}\right\|_{\diamond}
    &\leq\left\| \Xi_{\Sigma,\sigma^{1/2}} \circ \Phi\right\|_{\diamond} +\left\| \Xi_{\hat{\sigma}^{1/2},\Sigma} \circ \Phi\right\|_{\diamond} \nonumber \\
    &\quad +\left\|\Xi_{\hat{\sigma}^{1/2},\hat{\sigma}^{1/2}}\circ\left(\Phi-\hat{\Phi}\right)\right\|_{\diamond} \nonumber \\
    &\leq \|\Sigma\|_{\infty}\|\sigma^{1/2}\|_{\infty}\|\Phi\|_{\diamond}  \nonumber\\
    &\,\,+\|\hat{\sigma}^{1/2}\|_{\infty}\|\Sigma\|_{\infty}\|\Phi\|_{\diamond}\!+\! \|\hat{\sigma}^{1/2}\|_{\infty}^{2} \left\|\Phi \!-\!\hat{\Phi}\right\|_{\diamond} \nonumber \\
    &=\|\Sigma\|_{\infty}\left( \|\sigma^{1/2}\|_{\infty} + \|\hat{\sigma}^{1/2}\|_{\infty}\right)\|\Phi\|_{\diamond} \nonumber \\
    & \quad +\|\hat{\sigma}^{1/2}\|_{\infty}^{2} \left\|\Phi - \hat{\Phi} \right\|_{\diamond},
\end{align}
where the second inequality follows from Eq.~\eqref{eq:left_right_bound}. Next, notice that
\begin{align}
\label{eq:inner_difference_split}
    \Phi-\hat{\Phi}&=
    \mathcal{N}^{\dagger}\circ S_{\mathcal{N}(\sigma)^{-1/2}}
    -\hat{\mathcal{N}}^{\dagger}\circ S_{\hat{\mathcal{N}}(\hat{\sigma})^{-1/2}} \nonumber \\
    &= \left(\mathcal{N}^{\dagger}- \hat{\mathcal{N}}^{\dagger}\right)\circ S_{\mathcal{N}(\sigma)^{-1/2}}
    \nonumber \\
    & \,\,+\hat{\mathcal{N}}^{\dagger}\circ\left(S_{\mathcal{N}(\sigma)^{-1/2}}-S_{\hat{\mathcal{N}}(\hat{\sigma})^{-1/2}}\right).
\end{align}
Taking the diamond norm of this expression, and using the triangle inequality again yields
\begin{align}
    \left\|\Phi- \hat{\Phi} \right\|_{\diamond}&\leq
    \left\|\left( \mathcal{N}^{\dagger}\!-\!\hat{\mathcal{N}}^{\dagger} \right)\circ S_{\mathcal{N}(\sigma)^{-1/2}}\right\|_{\diamond} \nonumber \\
    & \quad +\left\|\hat{\mathcal{N}}^{\dagger}\!\circ\!\left(S_{\mathcal{N}(\sigma)^{-1/2}} \!-\! S_{\hat{\mathcal{N}}(\hat{\sigma})^{-1/2}}\right)\right\|_{\diamond}.
\end{align}
Substituting this bound into Eq.~\eqref{eq:diamond_triangle}, and noticing that $\|\hat{\sigma}^{1/2}\|_{\infty}^{2} = \|\hat{\sigma}\|_{\infty}$ since $\hat{\sigma}$ is positive yields Eq.~\eqref{eq:Dist_Petz_Bound}, which completes the proof.\qedhere
\end{proof}

\begin{lemma}\label{lem:lower_bound}
Let $\mathcal{N},\hat{\mathcal{N}}:\mathcal{L}(\mathcal{H}_A)\rightarrow \mathcal{L}(\mathcal{H}_B)$ be quantum channels and let $\sigma, \hat{\sigma}$ denote full-rank quantum states. Suppose $\mN(\sigma)$ and $\hat{\mN}(\hat{\sigma})$ are full-rank, then let $\mathcal{P}_\mathcal{N}^{\sigma}$ and $\mathcal{P}_{\hat{\mathcal{N}}}^{\hat\sigma}$ denote their corresponding Petz recovery maps. Then,
    \begin{align}\label{eq:Lower_BOund}
    \|\mathcal{P}_{\mathcal{N}}^\sigma\!-\!\mathcal{P}_{\hat{\mathcal{N}}}^{\hat{\sigma}} \|_\diamond \geq \max\Bigr\{&\norm{\sigma\!-\!\mathcal{P}_{\hat{\mathcal{N}}}^{\hat{\sigma}}(\mathcal{N}(\sigma))}_1,\nonumber \\
    & \quad \norm{\mathcal{P}_{\mathcal{N}}^{\sigma}(\hat{\mathcal{N}}(\hat{\sigma}))\!-\!\hat\sigma}_1\Bigr\}.
    \end{align}    
\end{lemma}
\begin{proof}
    Choosing a trivial auxiliary system and $\rho_B=\mathcal{N}(\sigma)$ in Eq.~\eqref{eq:def_DiamondDistance}, yields
    \begin{align}
        \|\mathcal{P}_{\mathcal{N}}^\sigma\!-\!\mathcal{P}_{\hat{\mathcal{N}}}^{\hat{\sigma}} \|_\diamond &\geq \norm{\mathcal{P}_{\mathcal{N}}^\sigma(\mathcal{N}(\sigma))\!-\!\mathcal{P}_{\hat{\mathcal{N}}}^{\hat{\sigma}}(\mathcal{N}(\sigma))}_1 \nonumber \\
        &\geq \norm{\sigma\!-\!\mathcal{P}_{\hat{\mathcal{N}}}^{\hat{\sigma}}(\mathcal{N}(\sigma))}_1,
    \end{align}
    where the last equality follows since the input state $\mathcal{N}(\sigma)$ is recovered exactly by the Petz recovery map. The lower bound in Eq.~\eqref{eq:Lower_BOund} follows since we can similarly choose $\rho_A=\hat{\mathcal{N}}(\hat{\sigma})$ and take the maximum over the two expressions. \qedhere
\end{proof}

\begin{corollary}\label{cor:bound_spec}
Let $\mathcal{N}:\mathcal{L}(\mathcal{H}_A)\rightarrow
\mathcal{L}(\mathcal{H}_B)$ be a quantum channel, and let $\sigma,\hat{\sigma}\in\mathcal{D}(\mathcal{H}_A)$
be full-rank states. Suppose $\mN(\sigma)$ is full-rank and suppose that
\begin{equation}
    \lambda_{\min}(\sigma)\geq\mu,\qquad\lambda_{\min}(\mathcal{N}(\sigma))\geq\nu,
\end{equation}
for some known constants $\mu,\nu>0$. If additionally
\begin{equation}
\label{eq:estimation_spectral_condition}
    \|\sigma-\hat{\sigma}\|_\infty\leq\frac{1}{2}\min\left\{\mu,\frac{\nu}{\|\mathcal{N}(\mathbb{1})\|_\infty}\right\},
\end{equation}
then
\begin{equation}
\label{eq:sigma_bound}
    \left\|\mathcal{P}_{\mathcal{N}}^{\sigma}-
    \mathcal{P}_{\mathcal{N}}^{\hat{\sigma}}\right\|_\diamond\leq C(\mu, \nu, \mathcal{N})\|\sigma-\hat{\sigma}\|_\infty,
\end{equation}
where
\begin{equation}
    C(\mu, \nu, \mathcal{N}):=\frac{\sqrt{2}\,\|\mathcal{N}(\mathbb{1})\|_\infty}{\nu\sqrt{\mu}}+\frac{4\|\mathcal{N}(\mathbb{1})\|_\infty^2}{\nu^2}.
\end{equation}
\end{corollary}
\begin{proof}
Let $\varepsilon:=\|\sigma-\hat{\sigma}\|_\infty.$ Using Weyl's eigenvalue perturbation inequality, see~\cite{Bhatia1997}, which in particular states that $|\lambda_{\min}(\hat{\sigma})-\lambda_{\min}(\sigma)|\leq \norm{\hat{\sigma}-\sigma}_\infty$, we obtain
\begin{align}
    \lambda_{\min}(\hat{\sigma})&\geq\lambda_{\min}(\sigma)-\|\sigma-\hat{\sigma}\|_\infty \geq\mu-\varepsilon\geq\frac{\mu}{2}.
\end{align}
Moreover, since $\mathcal{N}$ is positive, we have $\|\mathcal{N}(X)\|_\infty\leq
    \|\mathcal{N}(\mathbb{1})\|_\infty\|X\|_\infty$ for every Hermitian operator $X$. Therefore, it holds that
\begin{align}
    \lambda_{\min}(\mathcal{N}(\hat{\sigma}))
    &\geq\lambda_{\min}(\mathcal{N}(\sigma))-\|\mathcal{N}(\sigma-\hat{\sigma})\|_\infty \nonumber \\
    &\geq\nu-\|\mathcal{N}(\mathbb{1})\|_\infty\|\sigma-\hat{\sigma}\|_\infty \geq\tfrac{\nu}{2},
\end{align}
which implies
\begin{subequations}
\begin{align}
    \gamma&:=\min\left\{\lambda_{\min}(\sigma),\lambda_{\min}(\hat{\sigma})\right\}\geq\frac{\mu}{2}, \\
    \gamma'&:=\min\left\{\lambda_{\min}(\mathcal{N}(\sigma)),\lambda_{\min}(\mathcal{N}(\hat{\sigma}))\right\}\geq\frac{\nu}{2}.
\end{align}
\end{subequations}
We can now invoke Proposition~\ref{prop:main_bound}, which, here reduces to
\begin{align}\label{eq:alt_bound}
     &\|\mathcal{P}^\sigma_\mathcal{N} - \mathcal{P}^{\hat{\sigma}}_{\mN}\|_\diamond \nonumber \\
     &\leq  \|\Sigma\|_\infty (\|\sigma^{1/2}\|_\infty + \|\hat\sigma^{1/2}\|_\infty)\,\|\mathcal{N}^\dagger\circ S_{\mathcal{N}(\sigma)^{-1/2}}\|_\diamond\, \nonumber \\
     &\quad +\|\hat{\sigma}\|_\infty \|\mathcal{N}^\dagger \circ (S_{\mathcal{N}(\sigma)^{-1/2}} - S_{\mathcal{N}(\hat{\sigma})^{-1/2}})\|_\diamond\,,
\end{align}
where we recall that $\Sigma = \sigma^{\frac12}- \hat{\sigma}^{\frac12}$. We proceed to bound the first term in Eq.~\eqref{eq:alt_bound} first. To this end, notice that 
\begin{align}
    \|\mathcal{N}^{\dagger}
    \circ
    S_{\mathcal{N}(\sigma)^{-1/2}}\|_\diamond
    &=
    \|(\mathcal{N}^{\dagger}
    \circ
    S_{\mathcal{N}(\sigma)^{-1/2}})^\dagger(\mathbb{1})\|_\infty
     \nonumber \\
     &=\left\|\mathcal{N}(\sigma)^{-1/2}\mathcal{N}(\mathbb{1})\mathcal{N}(\sigma)^{-1/2}\right\|_\infty \nonumber \\
     &\leq\frac{\|\mathcal{N}(\mathbb{1})\|_\infty}{\nu}\,,
\end{align}
where we used the submultiplicativity of the operator norm. Moreover, using $\|\sigma^{1/2}\|_\infty\leq1$ and $\|\hat{\sigma}^{1/2}\|_\infty\leq1$ together with invoking Lemma~\ref{lem:infty_conti} to obtain $\norm{\Sigma}_\infty \leq \frac{\varepsilon}{2 \sqrt{\gamma}} \leq \frac{\varepsilon}{ \sqrt{2\mu}} $, yields the bound
\begin{align}\label{eq_first_term_bound}
     &\|\Sigma\|_\infty (\|\sigma^{1/2}\|_\infty + \|\hat\sigma^{1/2}\|_\infty)\,\|\mathcal{N}^\dagger\circ S_{\mathcal{N}(\sigma)^{-1/2}}\|_\diamond \nonumber \\
     &\leq \frac{\sqrt{2}\,\|\mathcal{N}(\mathbb{1})\|_\infty}
    {\nu\sqrt{\mu}}
    \|\sigma-\hat{\sigma}\|_\infty,
\end{align}
where we resubstituted $\varepsilon:=\|\sigma-\hat{\sigma}\|_\infty$. For the second contribution in Eq.~\eqref{eq:alt_bound}, we have the bound
\begin{align}\label{eq:second_term_bound}
    &\|\hat{\sigma}\|_\infty \|\mathcal{N}^\dagger \circ (S_{\mathcal{N}(\sigma)^{-1/2}} \!-\! S_{\mathcal{N}(\hat{\sigma})^{-1/2}})\|_\diamond \nonumber \\
    & \leq \|S_{\mathcal{N}(\sigma)^{-1/2}} \!-\! S_{\mathcal{N}(\hat{\sigma})^{-1/2}}\|_\diamond \norm{\mathcal{N}^\dagger}_\diamond
\end{align}
where we used submultiplicativity and that $\|\hat{\sigma}\|_\infty\leq1$ and  $\|\mathcal{N}^{\dagger}\|_\diamond=\|\mathcal{N}(\mathbb{1})\|_\infty$, which follows since  $\mathcal{N}^{\dagger}$ is completely positive and thus
\begin{align}
    \norm{\mN^\dagger}_\diamond&= \sup_{\rho} \Tr\left[ \idChan \otimes \mN^\dagger(\rho)\right]= \sup_\rho \Tr\left[\rho \mN(\mathbb{1})\right] \nonumber \\
    &= \norm{\mN(\mathbb{1})}_\infty.
\end{align}
Lastly, we obtain the bound
\begin{align}\label{eq:secondBOUnd}
& \left\|S_{\mathcal{N}(\sigma)^{-1/2}}-S_{\mathcal{N}(\hat{\sigma})^{-1/2}}\right\|_\diamond \nonumber \\
&\leq\left(\!\|\mathcal{N}(\sigma)^{-\tfrac{1}{2}}\|_\infty\!+\!
\|\mathcal{N}(\hat{\sigma})^{-\tfrac{1}{2}}\|_\infty\!\right)\left\|\mathcal{N}(\sigma)^{-\tfrac{1}{2}}\!-\!\mathcal{N}(\hat{\sigma})^{-\tfrac{1}{2}}\right\|_\infty\nonumber\\
&\leq\frac{1}{(\gamma')^2}\|\mathcal{N}(\sigma-\hat{\sigma})\|_\infty\leq\frac{\|\mathcal{N}(\mathbb{1})\|_\infty}{(\gamma')^2}\|\sigma-\hat{\sigma}\|_\infty.
\end{align}
Using $\gamma'\geq\nu/2$, we can therefore combine Eqs.~\eqref{eq:alt_bound},~ \eqref{eq_first_term_bound}, and~\eqref{eq:secondBOUnd}, which provides
\begin{align*}
    \left\|\mathcal{P}_{\mathcal{N}}^{\sigma}\!-\!
    \mathcal{P}_{\mathcal{N}}^{\hat{\sigma}}\right\|_\diamond\!\leq \!\left(\frac{\sqrt{2}\,\|\mathcal{N}(\mathbb{1})\|_\infty}{\nu\sqrt{\mu}}\!+\!\frac{4\|\mathcal{N}(\mathbb{1})\|_\infty^2}{\nu^2}\right)\|\sigma\!-\!\hat{\sigma}\|_\infty,
\end{align*}
proving the claim.\qedhere
\end{proof}

For a fixed reference state $\sigma$ and radius $\varepsilon>0$, let
\begin{equation}
    \mathcal{B}_{\varepsilon}(\sigma):=
    \left\{\hat{\sigma}\in\mathcal{D}(\mathcal{H}):\|\sigma-\hat{\sigma}\|_{1}\leq\varepsilon\right\}
\end{equation}
denote the trace-norm ball centered at $\sigma$. For any function $f(\sigma,\hat{\sigma})$, we define its normalized average over this ball as
\begin{equation}
    \left\langle f\right\rangle_{\varepsilon}:= \left(\int_{\mathcal{B}_{\varepsilon}(\sigma)}
        d\mu_{\rm HS}(\hat{\sigma})\right)^{-1}
    \displaystyle\int_{\mathcal{B}_{\varepsilon}(\sigma)} d\mu_{\rm HS}(\hat{\sigma})f(\sigma,\hat{\sigma}),
\end{equation}
where $\mu_{\rm HS}$ denotes the uniform, unitarily invariant volume measure induced by the Hilbert-Schmidt inner product $\langle A,B\rangle_{\mathrm{HS}}=\operatorname{Tr}(A^\dagger B)$.

\begin{figure}[ht]
    \centering
    \includegraphics[width=0.9\linewidth]{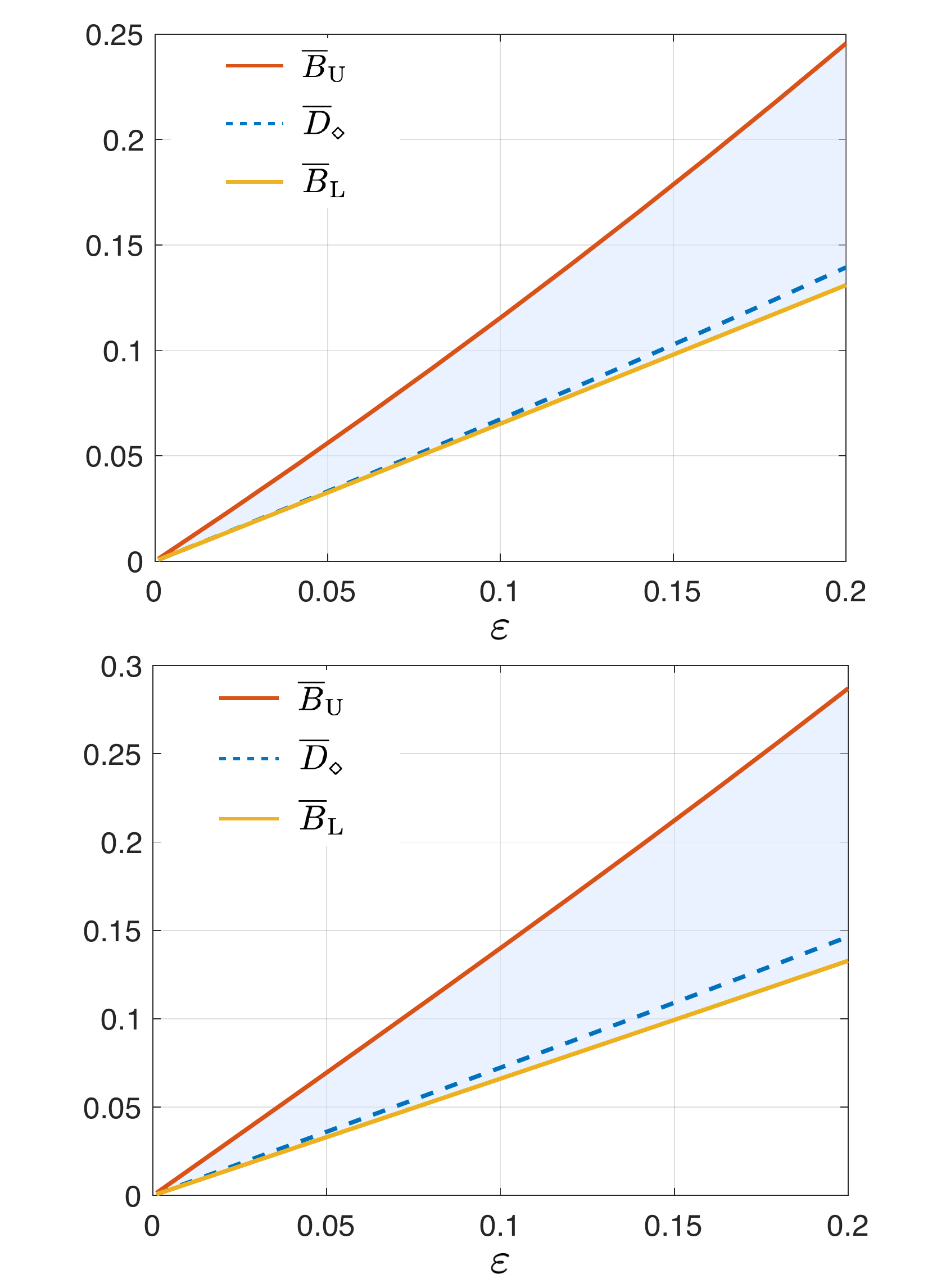}
    \caption{An example of the (averaged) diamond distance bounds between two Petz maps for the depolarizing channel with noise fixed at $p = 0.6$. We illustrate the averaged diamond norm (dashed blue) together with the averaged upper bound $\overline{B}_{\mathrm{U}}$ (orange) from Proposition~\ref{prop:main_bound}, and the averaged lower bound $\overline{B}_{\mathrm{L}}$ (yellow) from Lemma~\ref{lem:lower_bound}. Averaged here means that given a fixed state $\sigma$ chosen to be $\sigma = \mathbb{1}/2$ (upper panel) and $\sigma = \sigma(0.2,\pi/4, 0)$ (lower panel), where $\sigma(r,\theta, \phi) = \frac{1}{2}\begin{pmatrix}
        1 + r\cos{\theta} & r\sin{\theta}e^{-i\phi} \\
        r\sin{\theta}e^{i\phi} & 1 - r\cos{\theta}
    \end{pmatrix}$
    we consider states $\hat\sigma$ sampled uniformly with respect to the Hilbert-Schmidt measure inside the trace-norm ball $\|\sigma - \hat{\sigma}\|_1 \leq \varepsilon$. Note that the diamond distance can be computed using semidefinite programs~\cite{Watrous2009, Watrous2013}.
    }
    \label{fig:bound1}
\end{figure}

To illustrate the bounds, let us for convenience denote the upper bound from Proposition~\ref{prop:main_bound} by $B_{\rm U}$, and the lower bound from Lemma~\ref{lem:lower_bound} by $B_{\rm L}$. The
quantities displayed in Fig.~\ref{fig:bound1} are therefore
\begin{align}
    \overline{D}_{\diamond}(\varepsilon)
    &:=
    \left\langle
        \left\|
            \mathcal{P}_{\mathcal{N}}^{\sigma}
            -
            \mathcal{P}_{\mathcal{N}}^{\hat{\sigma}}
        \right\|_{\diamond}
    \right\rangle_{\varepsilon},
    \\
    \overline{B}_{\mathrm{U}}(\varepsilon)
    &:=
    \left\langle
        B_{\mathrm{U}}(\sigma,\hat{\sigma})
    \right\rangle_{\varepsilon},
    \\
    \overline{B}_{\mathrm{L}}(\varepsilon)
    &:=
    \left\langle
        B_{\mathrm{L}}(\sigma,\hat{\sigma})
    \right\rangle_{\varepsilon}.
\end{align}
Since the bounds hold for every $\hat{\sigma}$,
\begin{equation}
    B_{\mathrm{L}}(\sigma,\hat{\sigma})\leq
    \left\|\mathcal{P}_{\mathcal{N}}^{\sigma}-\mathcal{P}_{\mathcal{N}}^{\hat{\sigma}}\right\|_{\diamond}
    \leq B_{\mathrm{U}}(\sigma,\hat{\sigma}),
\end{equation}
their averages satisfy
\begin{equation}
    \overline{B}_{\mathrm{L}}(\varepsilon)\leq
    \overline{D}_{\diamond}(\varepsilon)\leq
    \overline{B}_{\mathrm{U}}(\varepsilon).
\end{equation}
The solid curves in Fig.~\ref{fig:bound1} are smoothed representations of the numerically evaluated averaged upper and lower bounds for the depolarizing channel. As expected, $\overline{D}_{\diamond}(\varepsilon)$ increases with the distance between the corresponding reference states. The region defined by the bounds closely follows the exact value of the diamond norm between Petz recovery maps associated with different reference states. 

In the upper panel, the average value of the diamond norm over randomly sampled states within the $\varepsilon-$ball remains below $\varepsilon = 0.2$, suggesting that $\|\mathcal{P}_{\mathcal{N}}^\sigma - \mathcal{P}_{\mathcal{N}}^{\hat{\sigma}} \|_\diamond \leq \|\sigma - \hat\sigma\|_1$ for the considered parameters. To provide a fair comparison with the bound, we average over randomly sampled perturbations in order to characterize the typical rather than the behavior for one state only. In the lower panel, we consider a reference state outside the principal Bloch axes. In contrast to the top panel, this state is not a fixed point of the depolarizing channel, providing a more generic scenario in which the validity of the bounds can be assessed. The average diamond norm again exhibits a nearly linear dependence on the distance between reference states, supporting the conclusion that the sensitivity of the Petz map to perturbations in the reference state remains under control. 

\section{Average fidelity for the IPR}
\label{ap:analytics}

In this section we derive the relation between the Haar average and the $2$-design average and show their relation for two classes of qubit channels. Recall that the $2$-design fidelity is defined as in Eq.~\eqref{eq:ipr_pauli_design_average}.

\begin{theorem}
Let $\mathcal{D}_p$ be the qubit depolarizing channel with $0\leq p <1$. For every pure input state $\psi$, the fidelity generated by the IPR using the maximally mixed initial reference state is independent of the input $\psi$ and satisfies $\overline{F}_k=F_k^{(2)}=F_k(\psi)=F_k$, where
\begin{equation}\label{eq:recurrence_rel}
F_{k+1}\!=\!F_k\left[\frac{(2\!-\!p)F_k}{2(1\!-\!p)F_k+p}\!+\!\frac{p(1\!-\!F_k)}{2-2(1\!-\!p)F_k-p}\right],
\end{equation}
with $F_1=1-p+\frac{p^2}{2}$. The sequence $\{F_k\}_{k\geq1}$ is strictly increasing and converges to unity. Moreover, its Haar average and $2$-design average coincide $\overline{F}_k=F_k^{(2)}=F_k(\psi)=a_k$. Furthermore,
\begin{equation}
\lim_{k\rightarrow\infty}\frac{1-F_{k+1}}{1-F_k}=\frac{p}{2-p},
\end{equation}
and hence the convergence is asymptotically geometric. For $p=0$, one has $F_k=1$ for all $k$.
\end{theorem}
\begin{proof}

Let $\psi$ denote an arbitrary but fixed qubit pure state, and let $\psi^\perp$ denote the orthogonal state to $\psi$. We begin by showing the recurrence relation in Eq.~\eqref{eq:recurrence_rel} for
\begin{align}
    F_k(\psi)=\Tr \left[\psi \left(\mathcal{P}_{\mD_p}^{\sigma_{k-1}}\circ \ldots \circ\mathcal{P}_{\mD_p}^{\sigma_{0}}\circ\mD_p\right)(\psi)\right].
\end{align}
Recall that the first step ($k=1$) in the IPR (using the maximally mixed state as the reference state) simply applies $\mD_p$ again. Thus, a straightforward calculation reveals that the first step of the IPR outputs the state
\begin{align}
    \rho_1  &=\left(\mD_p\circ\mD_p\right)(\psi)=\mD_p\left((1-p)\psi+p\frac{\mathbb{1}}{2}\right) \nonumber\\
    &= \left(1-p+\frac{p^2}{2}\right) \psi+p\left(1-\frac{p}{2}\right) \psi^\perp. 
\end{align}
Hence, we find that
\begin{align}
    F_1(\psi)=\Tr \left[\psi \rho_1\right]=1-p+\frac{p^2}{2},
\end{align}
which is independent of $\psi$, and we can thus also write $\rho_1=F_1\psi+(1-F_1)\psi^\perp$.
We now proceed by induction. Suppose that, for some $k\geq1$, the recovered state is given by
\begin{equation}
    \rho_k=F_k\psi+(1-F_k)\psi^\perp.
\end{equation}
Then, 
\begin{align}\label{eq:depolarizingNoisyReference}
    \mD_p(\rho_k)
    &=\left[  \left(1-\frac{p}{2}\right)F_k +\frac{p}{2}(1-F_k) \right]\psi \nonumber \\
    &\quad +\left[\frac{p}{2}F_k+\left(1-\frac{p}{2}\right)(1-F_k) \right]\psi^\perp \nonumber \\
    &=: \alpha_k\psi+(1-\alpha_k)\psi^\perp,
\end{align}
where we introduced $\alpha_k:=\left(1-\frac{p}{2}\right)F_k +\frac{p}{2}(1-F_k)$ for ease of notation. Notice that the latter expression is diagonal in the basis $\{\psi,\psi^\perp\}$ and thus
\begin{align}\label{eq:depolarizingRescaledState}
    \mD_p(\rho_k)^{-1/2}\rho_k\mD_p(\rho_k)^{-1/2}&= \frac{F_k}{\alpha_k}\psi+ \frac{1-F_k}{1-\alpha_k}\psi^\perp.
\end{align}
For arbitrary coefficients $b$ and $c$, the action of the
depolarizing channel on an operator diagonal in this basis is
\begin{align*}
    \mD_p(b\psi\!+\!c\psi^\perp)\!=\!\left[\left(1\!-\!\frac{p}{2}\right)b\!+\!\frac{p}{2}c\right]\psi \!+\!\left[\frac{p}{2}b\!+\!\left(1\!-\!\frac{p}{2}\right)c\right]\psi^\perp,
\end{align*}
from which it follows that
\begin{align}
    &\mD_p\!\left(\mD_p(\rho_k)^{- \tfrac{1}{2}} \rho_k\mD_p(\rho_k)^{- \tfrac{1}{2}}\right) \! =\!\left[\left(1\!-\!\frac{p}{2}\right)
        \frac{F_k}{\alpha_k}\!+\!\frac{p}{2}\frac{1\!-\!F_k}{1\!-\!\alpha_k}\right]\psi \nonumber \\
        &\qquad \qquad \qquad+\left[\frac{p}{2}\frac{F_k}{\alpha_k}+\left(1-\frac{p}{2}\right)\frac{1-F_k}{1-\alpha_k}\right]\psi^\perp.
    \label{eq:depolarizingMiddleAction}
\end{align}
Since $\mD_p$ is self-dual, sandwiching this expression between $\sqrt{\rho_k}=\sqrt{F_k}\,\psi+\sqrt{1-F_k}\,\psi^\perp$ gives
\begin{align}
    \rho_{k+1}&=\rho_k^{\tfrac{1}{2}}\mD_p\!\left(\mD_p(\rho_k)^{- \tfrac{1}{2}} \rho_k\mD_p(\rho_k)^{- \tfrac{1}{2}}\right)\rho_k^{\tfrac{1}{2}} \nonumber \\
    &= F_k\left[\left(1-\frac{p}{2}\right)\frac{F_k}{\alpha_k}+\frac{p}{2}\frac{1-F_k}{1-\alpha_k}\right] \psi \nonumber \\
    & \, + \left(1-F_k\left[\left(1-\frac{p}{2}\right)\frac{F_k}{\alpha_k}+\frac{p}{2}\frac{1-F_k}{1-\alpha_k}\right]\right)\psi^\perp \nonumber \\
    &=: F_{k+1} \psi + (1-F_{k+1}) \psi^\perp.
\end{align}
Moreover, this implies that
\begin{equation}
    F_k(\psi)={\rm Tr}[\psi\rho_k]=F_k. \quad \forall k\geq 1,
\end{equation}
where the coefficient $F_k$ is independent of the input state $\psi$ and solely depends on $p$ (and $k$), which proves the recurrence relation in Eq.~\eqref{eq:recurrence_rel} by induction. From this, we immediately find that
\begin{equation}
    \overline F_k =\int \,d\mu(\psi) \,{\rm Tr}[\psi\rho_k]  = F_k,
\end{equation}
and analogously
\begin{align}
    \overline F_k^{(2)}=
    \frac{1}{6}\sum_{\vec r\in\{\pm e_x,\pm e_y,\pm e_z\}}F_{k}(\psi_{\vec{r}})=F_k.
\end{align}
To show convergence, we compute
\begin{equation}
F_{k+1}\!-\!F_k=\frac{2p(1\!-\!p)F_k(1\!-\!F_k)(2F_k-1)}{\left[2(1\!-\!p)F_k\!+\!p\right]\left[2\!-\!2(1\!-\!p)F_k\!-\!p\right]}.
\end{equation}
Since $F_1 \in\left(\frac12,1\right)$, the right-hand side is strictly positive whenever $F_k\in(1/2,1)$. Hence $\{F_k\}$ is strictly increasing and bounded above by one, and therefore convergent. The only fixed point of the
recurrence in $(1/2,1]$ is $F=1$, implying $\lim_{k\rightarrow\infty}F_k=1$. Expanding the recurrence around this fixed point yields
\begin{equation}
1-F_{k+1}=\frac{p}{2-p}(1-F_k)+\mathcal{O}\!\left((1-F_k)^2\right).
\end{equation}
Thus, it follows that
\begin{equation}
\lim_{k\rightarrow\infty}\frac{1-F_{k+1}}{1-F_k}=\frac{p}{2-p}.
\end{equation}
The case $p=0$ is immediate, since the channel is the identity and $F_k=1$ for every $k$.\qedhere
\end{proof}

To establish a relation between the Haar average for the single-axis Pauli channel and the $2$-design average, we first show that the IPR fidelity inherits the axial symmetry of the channel. In other words, for any pure input state with $|\vec r| = 1$, the fidelity $F_k(\vec r)$ depends solely on the direction in which the noise channel acts.
\begin{lemma}
\label{lem:ipr_axial_symmetry}
Let $\psi (\vec{r})$ be a pure qubit state, with Bloch vector $\vec{r}$ and let $\mathcal{N}$ be a quantum channel, covariant under rotations around an axis $\alpha$, i.e., 
\begin{align}\label{eq:cov}
    \mN\circ \mU_{\alpha,\theta}=\mU_{\alpha,\theta} \circ \mN
\end{align} 
where $\alpha \in\{ x, y, z\}$. Assume that the reference sequence is initialized with an axially symmetric full-rank state. Then, for every $k\geq1$, the fidelity of the IPR 
\begin{equation}
    F_k(\vec r)= {\rm Tr}\!\left[\psi(\vec r)\rho_k(\vec r)\right]
\end{equation}
is axially symmetric around $\alpha$. Equivalently, there exists a function $ g_k:[-1,1]\to\mathbb{R}$ such that $F_k(\vec r)=g_k(r_\alpha)$, where $r_\alpha$ denotes the $\alpha$-component of the Bloch vector $\vec{r}$.
\end{lemma}
\begin{proof}
Let $\mathcal U_{\alpha, \theta}(X)=U_{\alpha, \theta} XU_{\alpha, \theta}^\dagger $, where
\begin{equation}
    U_{\alpha, \theta}=e^{-i\theta P_\alpha/2}\,,
\end{equation}
for $P_\alpha$ a Pauli operator. Let $\sigma$ be a full-rank state and define $\sigma_{\theta, \alpha}:= \mathcal{U}_{\alpha, \theta}(\sigma)$. Since $\mathcal N$ is covariant under rotations around the $\alpha$-axis, namely, we have $\mathcal N(\sigma_{\theta, \alpha}) =\mathcal N(\mathcal U_{ \alpha, \theta}(\sigma))=\mathcal U_{ \alpha, \theta}(\mathcal N(\sigma))$, for every $\theta$. We can therefore write 
\begin{subequations}
\begin{align}
    \sigma_{\theta, \alpha}^{1/2}&=U_{ \alpha, \theta} \sigma^{1/2} U_{ \alpha, \theta}^\dagger, \\
    \mathcal N(\sigma_{\theta, \alpha})^{-1/2}&=U_{ \alpha, \theta} \mathcal N(\sigma)^{-1/2} U_{ \alpha, \theta}^\dagger .
\end{align}
\end{subequations}
Moreover, from Eq.~\eqref{eq:cov}, it follows, by taking the Hilbert-Schmidt adjoint, that $\mathcal N^\dagger\circ \mathcal U_{ \alpha, \theta}^\dagger=\mathcal U_{ \alpha, \theta}^\dagger \circ \mathcal N^\dagger$.
Equivalently, this implies that $\mathcal N^\dagger$ is also covariant with respect to the same unitary symmetry. We now show that the Petz map transforms covariantly. For any operator $X$, we have
\begin{widetext}
\begin{align}
\label{eq:cov_petz}
    \mathcal P_{\mathcal N}^{\sigma_{\theta, \alpha}}
    \bigl(\mathcal U_{ \alpha, \theta}(X)\bigr) &=\sigma_{\theta, \alpha}^{1/2}
    \mathcal N^\dagger \left[  \mathcal N(\sigma_{\theta, \alpha})^{-1/2}
        \mathcal U_{ \alpha, \theta}(X)
        \mathcal N(\sigma_{\theta, \alpha})^{-1/2}
    \right]\sigma_{\theta, \alpha}^{1/2} \nonumber \\
    &=\sigma_{\theta, \alpha}^{1/2} 
    \mathcal N^\dagger
    \left[
        U_{ \alpha, \theta} \mathcal N(\sigma)^{-1/2} U_{ \alpha, \theta}^\dagger
        U_{ \alpha, \theta} X U_{ \alpha, \theta}^\dagger
        U_{ \alpha, \theta} \mathcal N(\sigma)^{-1/2} U_{ \alpha, \theta}^\dagger
    \right]\sigma_{\theta, \alpha}^{1/2}\nonumber\\
    &=U_{ \alpha, \theta}\sigma^{1/2}U_{ \alpha, \theta}^\dagger
    \mathcal N^\dagger
    \left[
        U_{ \alpha, \theta}
        \mathcal N(\sigma)^{-1/2}
        X
        \mathcal N(\sigma)^{-1/2}U_{ \alpha, \theta}^\dagger\right] U_{ \alpha, \theta}\sigma^{1/2}U_{ \alpha, \theta}^\dagger\nonumber\\
    &=U_{ \alpha, \theta}\sigma^{1/2}
    \mathcal N^\dagger
    \left[\mathcal N(\sigma)^{-1/2}X\mathcal N(\sigma)^{-1/2}\right]\sigma^{1/2}U_{ \alpha, \theta}^\dagger =\mathcal U_{ \alpha, \theta}
    \left(\mathcal P_{\mathcal N}^{\sigma}(X)\right).
\end{align}
\end{widetext}
Note that the reference state on the right-hand side is the original reference state $\sigma$. If $\sigma$ is itself invariant under the rotation, the relation reduces to the usual covariance condition for the fixed Petz map,
\begin{equation}
\mathcal P_{\mathcal N}^{\sigma}
\circ\mathcal U_{\alpha,\theta}
=
\mathcal U_{\alpha,\theta}
\circ\mathcal P_{\mathcal N}^{\sigma}.
\end{equation}
We now apply this covariance property to the iterative protocol. The
unitary $U_{\alpha, \theta}$ induces the corresponding rotation of the Bloch vector:
\begin{equation}
\psi\bigl(R_\alpha(\theta)\vec r\bigr)=\mathcal U_{\alpha,\theta}\bigl(\psi(\vec r)\bigr).
\end{equation}
Hence,
\begin{align}
    \rho(R_\alpha(\theta)\vec r)
    &=
    \mathcal N\bigl(\psi(R_\alpha(\theta)\vec r)\bigr)=
    \mathcal N\bigl(\mathcal U_{ \alpha, \theta}(\psi(\vec r))\bigr) \nonumber \\
    &=
    \mathcal U_{ \alpha, \theta}
    \bigl(\mathcal N(\psi(\vec r))\bigr)=
    \mathcal U_{ \alpha, \theta}(\rho(\vec r)).
\end{align}
The initial reference state, taken to be the maximally mixed state is invariant and satisfies
\begin{equation}
    \sigma_0(R_\alpha(\theta)\vec r)
    = \sigma_0 = 
    \mathcal U_{ \alpha, \theta}(\sigma_0(\vec r)).
\end{equation}
Thus, Eq.~\eqref{eq:cov_petz} holds for $k =1$. We proceed by induction. Suppose that the covariance property holds for some $k$.

Then, we obtain 
\begin{align}
\rho_{k+1}\bigl(R_\alpha(\theta)\vec r\bigr)&=
\mathcal P_{\mathcal N}^{\sigma_k(R_\alpha(\theta)\vec r)}\left(\rho_k(R_\alpha(\theta)\vec r)\right) \nonumber \\
&=\mathcal P_{\mathcal N}^{
\mathcal U_{\alpha,\theta}(\sigma_k(\vec r))}
\left(
\mathcal U_{\alpha,\theta}(\rho_k(\vec r))
\right) \nonumber \\
&=
\mathcal U_{\alpha,\theta}
\left(
\mathcal P_{\mathcal N}^{\sigma_k(\vec r)}
\bigl(\rho_k(\vec r)\bigr)
\right)\nonumber \\
&=\mathcal U_{\alpha,\theta}
\bigl(\rho_{k+1}(\vec r)\bigr).
\end{align}
Since the next reference state is precisely the recovered state, we have
\begin{align}
\sigma_{k+1}\bigl(R_\alpha(\theta)\vec r\bigr)&=
\rho_{k+1}\bigl(R_\alpha(\theta)\vec r\bigr)=\mathcal U_{\alpha,\theta}
\bigl(\rho_{k+1}(\vec r)\bigr) \nonumber \\
&=\mathcal U_{\alpha,\theta}\bigl(\sigma_{k+1}(\vec r)\bigr).
\end{align}
This completes the induction. Therefore both sequences $\{\rho_k\}$ and $\{\sigma_k\}$ transform covariantly for all $k\geq1$.

Finally, we consider the fidelity. Using the covariance of $\rho_k$ and the transformation law for $\psi$, we find
\begin{align}
F_k\bigl(R_\alpha(\theta)\vec r\bigr)
&={\rm Tr}\left[\psi\bigl(R_\alpha(\theta)\vec r\bigr)\rho_k\bigl(R_\alpha(\theta)\vec r\bigr)\right] \nonumber \\
&={\rm Tr}\left[\mathcal U_{\alpha,\theta}\bigl(\psi(\vec r)\bigr)\mathcal U_{\alpha,\theta}\bigl(\rho_k(\vec r)\bigr)\right] \nonumber \\
&={\rm Tr}\left[\psi(\vec r)\rho_k(\vec r)\right]
=F_k(\vec r),
\end{align}
where we used cyclicity of the trace. For completeness, we now show explicitly why rotational invariance implies that the fidelity depends only on the component of the Bloch vector along the
symmetry axis. We consider the case $\alpha=z$. Let
$\vec r=(r_x,r_y,r_z)$ be a unit Bloch vector. The transverse components satisfy
\begin{equation}
r_x^2+r_y^2=1-r_z^2.
\end{equation}
Hence there exists an angle $\phi$ such that
\begin{equation}
\vec r=\left(\sqrt{1-r_z^2}\cos\phi,\sqrt{1-r_z^2}\sin\phi,r_z\right).
\end{equation}
Equivalently,
\begin{equation}
\vec r=R_z(\phi)\vec r_0(r_z),
\end{equation}
where
\begin{equation}
\vec r_0(r_z):=\left(\sqrt{1-r_z^2},0,r_z\right).
\end{equation}
Using the rotational invariance of the fidelity,
\begin{equation}
F_k\bigl(R_z(\phi)\vec r_0(r_z)\bigr)=F_k\bigl(\vec r_0(r_z)\bigr),
\end{equation}
we obtain
\begin{equation}
F_k(\vec r)=F_k\bigl(\vec r_0(r_z)\bigr).
\end{equation}
We may therefore define
\begin{equation}
g_k(r_z):=F_k\bigl(\vec r_0(r_z)\bigr),\qquad r_z\in[-1,1].
\end{equation}
The same argument applies to rotations around the $x$- and $y$-axis after replacing $r_z$ by the corresponding component $r_\alpha$.
\end{proof}

For an axially covariant channel, Lemma~\ref{lem:ipr_axial_symmetry}
reduces the Haar integral to one dimension and gives the Pauli $2$-design average a useful quadrature interpretation. 

\begin{lemma}
\label{lem:simpson}
    Let $\mathcal N$ be covariant under rotations around the $\alpha$-axis, and assume that the fidelity after the $k-th$ IPR step depends on the input Bloch vector only through its axial component $F_k(\vec r)=g_k(r_\alpha)$ (see Lemma~\ref{lem:ipr_axial_symmetry}). Thus, the Pauli $2$-design average fidelity $\overline F_k^{(2)}$ coincides exactly with the Simpson quadrature associated with the one-dimensional Haar integral~\cite{Atkinson1989}. Therefore, the difference between the Haar and Pauli $2$-design averages is precisely the corresponding Simpson quadrature error.
\end{lemma}
\begin{proof}
Without loss of generality, take $\alpha=z$. The Haar-averaged fidelity over all pure qubit states is 
\begin{equation}
\overline F_k = \int_{\mathbb S^2} d\mu(\vec r)\, F_k(\vec r)\,, 
\end{equation}
where $d\mu$ denotes the normalized uniform measure on the Bloch sphere denoted by $\mathbb S^2$. Parametrizing a unit Bloch vector as $\vec r=(\sin\theta\cos\phi,\sin\theta\sin\phi,\cos\theta)$, the normalized Haar measure becomes 
\begin{equation}
d\mu = \frac{1}{4\pi}\sin{\theta}\, d\theta\,d\phi.
\end{equation}
Axial symmetry implies that the fidelity depends on $\vec{r}$ only through its component along the symmetry axis, 
\begin{equation}
    r_z = \cos{\theta}, \qquad F_k(\vec r(\theta,\phi))=g_k(\cos\theta).
\end{equation}
Therefore,  
\begin{align}
    \overline F_k
    &=\frac{1}{4\pi}
    \int_0^{2\pi}\mathrm d\phi
    \int_0^\pi\mathrm d\theta\,
    \sin\theta\,g_k(\cos\theta)
    \nonumber\\
    &=\frac{1}{2}\int_{-1}^{1}g_k(u)\,\mathrm du\,
\end{align}
where we introduced the variable $u = \cos{\theta}$. The integral above can be directly connected with the Simpson quadrature for an integral over $[-1,1]$ \cite{Atkinson1989} \begin{equation}
\label{eq:simpson}
\int_{-1}^{1} du\,g_k(u)\approx\frac{1}{3}\left[g_k(-1)+4g_k(0)+g_k(1)\right].
\end{equation}
We now consider the six states forming the Pauli $2$-design, whose average fidelity is
\begin{equation}
    F_k^{(2)}:=
    \frac{1}{6}\sum_{\vec r\in\mathcal S_{\mathrm P}}F_k(\vec r),\quad \mathcal S_{\mathrm P} :=\{\pm\vec e_x,\pm\vec e_y,\pm\vec e_z\}\,.
\end{equation}
The two eigenstates along the $z$-axis have $r_z = \pm 1$, while the four eigenstates along the $x$- and $y$-axis have $r_z = 0$. Their uniform average is therefore 
\begin{equation}
     F_k^{(2)}= \frac{1}{6}\left[g_k(-1)+4g_k(0)+g_k(1)\right]\,.
\end{equation}
Observe that the expression above is exactly Eq.~\eqref{eq:simpson} multiplied by $1/2$. Thus, we obtain 
\begin{equation}
    \overline F_k \approx \frac{1}{6}\left[g_k(-1)+4g_k(0)+g_k(1)\right] = F_k^{(2)}. 
\end{equation}
Hence, the Pauli $2$-design average coincides exactly with the Simpson quadrature associated with the one-dimensional Haar integral. This does not, in general, imply exact equality with the Haar average. It identifies $F_k^{(2)}$ as its Simpson approximation.
\end{proof}

\begin{figure*}[ht]
    \centering
    \includegraphics[width=0.45\linewidth]{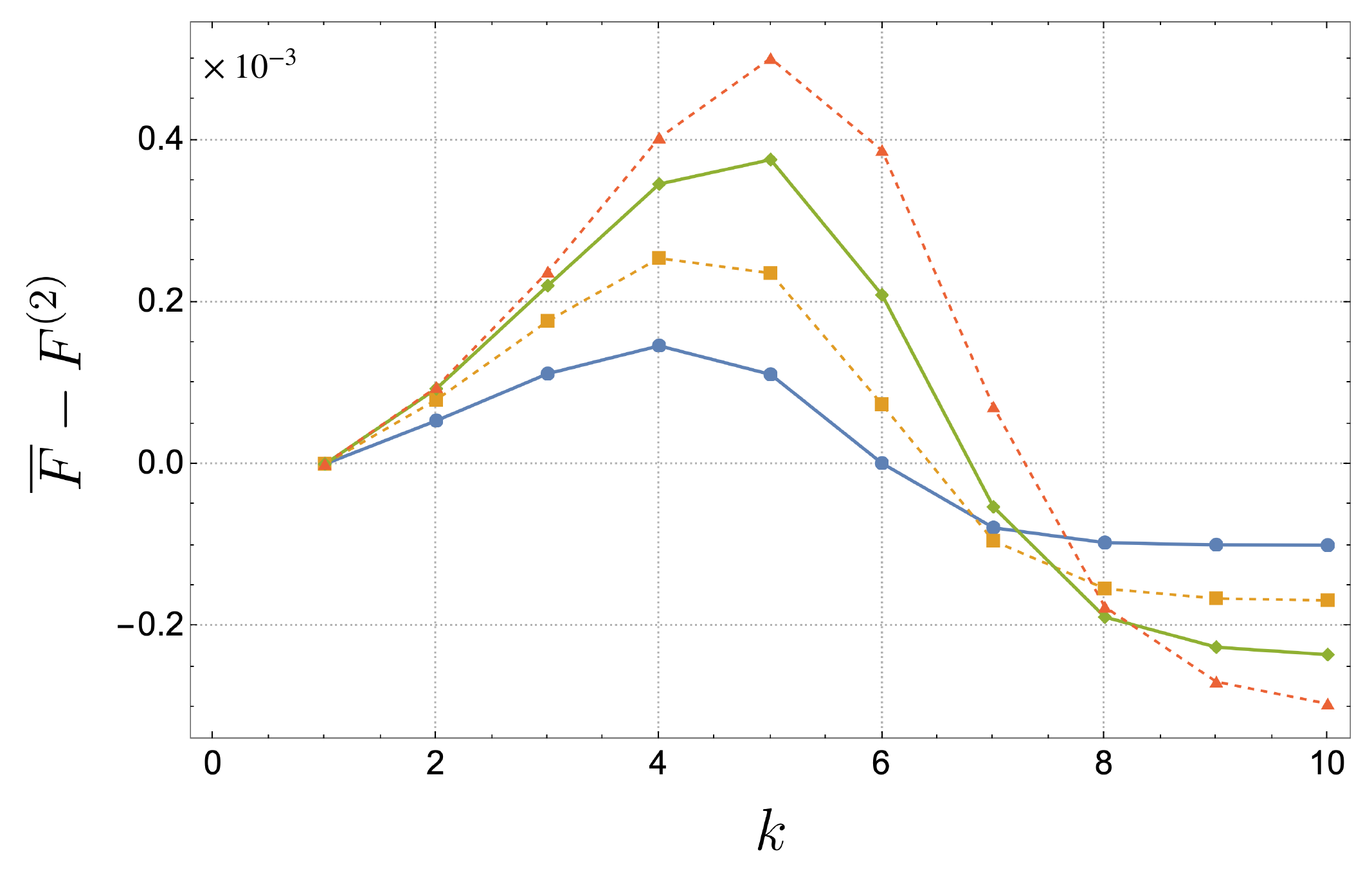}
    \includegraphics[width=0.45\linewidth]{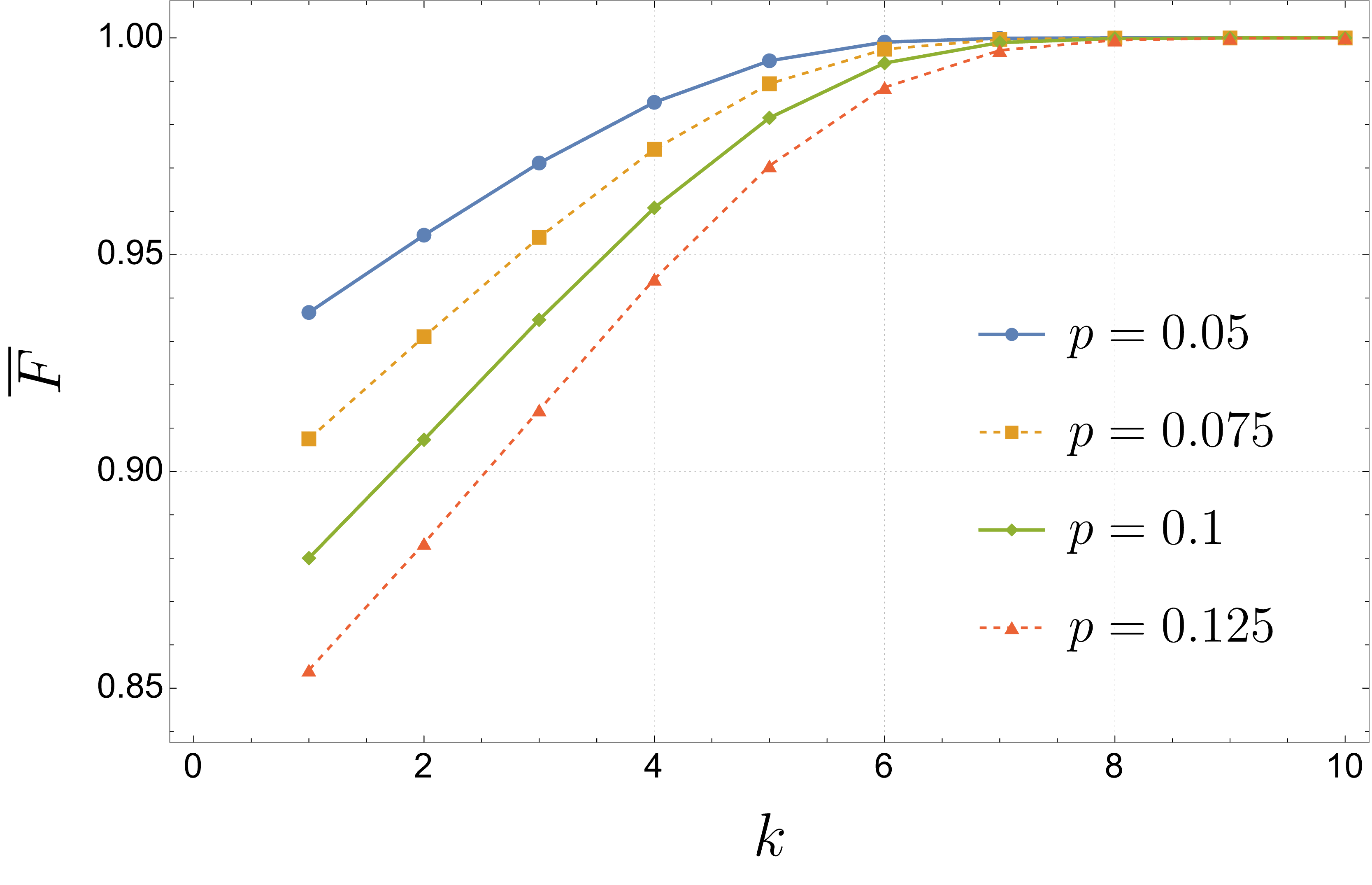}
    \caption{Numerical comparison between the Haar and Pauli $2$-design averages for dephasing along the $x$-axis. For each noise strength $p$ and iteration number $k$, the Haar average was obtained by numerically evaluating the integral in Eq.~\eqref{eq:haar_avg}. Numerical convergence was verified by repeating the quadrature with progressively tighter integration tolerances. Left: difference between the two averages as a function of IPR iterations for different noise strengths. The discrepancy decreases as the noise strength is reduced and remains of the order of $10^{-4}$ for the considered parameter range. Right: Corresponding Haar-averaged fidelity as a function of the number of iterations for the same noise strengths. Because the Haar and Pauli $2$-design curves overlap at the displayed scale, only the Haar-averaged curves are shown.}
    \label{fig:dif_avgfidelity}
\end{figure*}

To quantify the accuracy of the Pauli $2$-design approximation for this class of channels, we compare it numerically with the corresponding Haar average. Figure~\ref{fig:dif_avgfidelity} left panel shows the difference between the two averages for a single-axis Pauli channel, while the right panel displays the corresponding Haar-averaged fidelity as a function of the number of iterations of the IPR. Throughout the considered parameter range, the discrepancy remains of the order of $10^{-4}$, confirming that the Pauli $2$-design provides an accurate approximation to the Haar average for the studied parameter range. Over the finite iteration range shown, the Haar and
Pauli $2$-design averages remain close. This finite-depth agreement does not determine their asymptotic behavior: the $2$-design average converges to unity, whereas the Haar average develops a small but nonvanishing asymptotic deficit. 

We now derive an analytical expression for the former under the IPR for this class of channels. 

\begin{theorem}
Let $\mathcal{E}_{\alpha,p}$ be a single-axis Pauli channel with $0\leq p<1/2$. The Pauli $2$-design average fidelity generated by the IPR is 
\begin{equation}
    F^{(2)}_k=\frac{1+2a_k}{3},
\end{equation}
where
\begin{equation}
a_{k+1}\!=\!a_k\left[\frac{(1\!-\!p)a_k}{a_k\!+\!p\!-\!2a_kp}\!+\!\frac{p(1\!-\!a_k)}{1\!-\!a_k\!-\!p\!+\!2a_kp}\right],
\end{equation}
with $a_1=1-2p(1-p)$. For $0<p<1/2$, the sequence $\{a_k\}_{k\geq1}$ is strictly increasing and converges to unity. Thus,
\begin{equation}
    \lim_{k\rightarrow\infty}F^{(2)}_k=1.
\end{equation}
Moreover,
\begin{equation}
    \lim_{k\rightarrow\infty}\frac{1-F^{(2)}_{k+1}}{1-F^{(2)}_k}=\frac{p}{1-p},
\end{equation}
so the convergence is asymptotically geometric. For $p=0$, one has $F^{(2)}_k=1$ for all $k$.
\end{theorem}
\begin{proof}
The axial eigenstates $\psi_\alpha$ and $\psi_\alpha^\perp$ are fixed points of $\mathcal{E}_{\alpha,p}$. They are therefore also fixed by every step of the iterative Petz protocol, and their recovery fidelity is equal to one for all $k$. Therefore, we exclude them from the nontrivial recursive construction. In particular, the pure axial states are never used as reference states. We now consider a transverse direction $i\neq\alpha$. Given that the channel is unital and self-dual, after the first Petz step we obtain
\begin{equation}
    \rho_{i,1}=a_1\psi_i+(1-a_1)\psi_i^\perp,
\end{equation}
where $a_1=1-2p(1-p)$. We again proceed by induction and assume that for some $k$ it holds that
\begin{equation}
    \rho_{i,k}=a_k\psi_i+(1-a_k)\psi_i^\perp\,,
\end{equation}
we get
\begin{equation}
\label{eq:noisyReferenceSingleAxis}
    \mathcal{E}_{\alpha,p}(\rho_{i,k})=\alpha_k\psi_i+(1-\alpha_k)\psi_i^\perp,
\end{equation}
with $\alpha_k:=(1-p)a_k+p(1-a_k)$. Since all the operators involved are diagonal in the basis $\{\psi_i,\psi_i^\perp\}$,
\begin{align}
    &\mathcal{E}_{\alpha,p}(\rho_{i,k})^{-\tfrac{1}{2}}
    \rho_{i,k}
    \mathcal{E}_{\alpha,p}(\rho_{i,k})^{-\tfrac{1}{2}} \!= \! \frac{a_k}{\alpha_k}\psi_i\!+\!\frac{1\!-\!a_k}{1\!-\!\alpha_k}\psi_i^\perp.
    \label{eq:singleAxisRescaledState}
\end{align}
Applying the channel once more gives
\begin{align}\label{eq:singleAxisChannelMiddle}
   & \mathcal{E}_{\alpha,p}
    \left(\mathcal{E}_{\alpha,p}(\rho_{i,k})^{-\tfrac{1}{2}}\rho_{i,k}\mathcal{E}_{\alpha,p}(\rho_{i,k})^{-\tfrac{1}{2}}\right)  \\
    &=\left[(1\!-\!p)\frac{a_k}{\alpha_k}\!+\!p\frac{1\!-\!a_k}{1\!-\!\alpha_k}\right]\psi_i 
     \!+\!\left[p\frac{a_k}{\alpha_k}+(1\!-\!p)\frac{1\!-\!a_k}{1\!-\!\alpha_k}\right]\psi_i^\perp \nonumber.
\end{align}
Here we used the self-duality of the channel $\mathcal{E}_{\alpha,p}$. Finally, sandwiching Eq.~\eqref{eq:singleAxisChannelMiddle} between the square roots of
\begin{equation}
    \rho_{i,k}^{1/2}=\sqrt{a_k}\,\psi_i+\sqrt{1-a_k}\,\psi_i^\perp
\end{equation}
yields
\begin{equation}
    \rho_{i,k+1}=a_{k+1}\psi_i+(1-a_{k+1})\psi_i^\perp,
\end{equation}
where
\begin{equation}
    a_{k+1}=a_k\left[(1-p)\frac{a_k}{\alpha_k}+p\frac{1-a_k}{1-\alpha_k} \right].
\end{equation}
Substituting $\alpha_k$ gives precisely Eq.~\eqref{eq:singleAxisRecurrence}, completing the induction. For a pure input state $\psi_i$, the fidelity is
\begin{equation}
    F(\psi_i,\rho_{i,k})
    ={\rm Tr} [\psi_i\rho_{i,k}]= a_k.
\end{equation}
The same result holds for the corresponding orthogonal input $\psi_i^\perp$. Thus, among the six Pauli eigenstates, the two axial states have fidelity one and the four transverse states have fidelity
$a_k$. It follows that
\begin{equation}
    F_k^{(2)}=\frac{1}{6}\left(2+4a_k\right)=
    \frac{1+2a_k}{3}.
\end{equation}
Using the previously established result in Lemma~\ref{lem:simpson}, we complete the proof. To show convergence, we compute 
\begin{equation}
a_{k+1}\!-\!a_k=\frac{p(1\!-\!2p)a_k(1\!-\!a_k)(2a_k\!-\!1)}{\left(a_k\!+\!p\!-\!2a_kp\right)\left(1\!-\!a_k\!-\!p\!+\!2a_kp\right)}.
\end{equation}
Moreover, $a_1\in\left(\frac12,1\right)$. Thus $a_{k+1}>a_k$ whenever $a_k\in(1/2,1)$. The sequence is therefore strictly increasing and bounded above by one. Its only possible fixed point in $(1/2,1]$ is $a=1$, and hence $\lim_{k\rightarrow\infty}a_k=1$. It follows from $F_k^{(2)}=(1+2a_k)/3$ that $\lim_{k\rightarrow\infty}F_k^{(2)}=1$.

Expanding the recurrence around $a=1$ gives
\begin{equation}
1-a_{k+1}=\frac{p}{1-p}(1-a_k)+\mathcal{O}\!\left((1-a_k)^2\right).
\end{equation}
Because $1-F_k^{(2)}=\frac{2}{3}(1-a_k)$, this implies
\begin{equation}
\lim_{k\rightarrow\infty}
\frac{1-F_{k+1}^{(2)}}{1-F_k^{(2)}}=\frac{p}{1-p}.
\end{equation}
For $p=0$, the fidelity is identically one. At the boundary
$p=1/2$, one instead has $a_k=1/2$ and $F_k^{(2)}=2/3$ for every $k$.
\end{proof}

\section{Finite-sample implementation of the Petz recovery map}
\label{ap:implement}
The implementation of the Petz recovery map has been studied in ~\cite{Wilde22, Prabha24}, with specific approaches based on the quantum singular value transform being identified. Naturally, these implementations require knowledge of both the error channel and the reference state. In particular, the construction in ~\cite{Wilde22} requires access to either a purification of the reference state or its classical description. When only physical copies of the reference state are available, one may perform state tomography to allow for such an implementation.

In this section, using the bounds derived in
Appendix~\ref{ap:diamond_bounds}, we derive a sufficient
number of copies for implementing the Petz map to a prescribed accuracy. We do not restrict the analysis to a particular tomographic estimator or measurement scheme. Instead, for a chosen tomographic procedure, we denote by $N_{\mathrm{tomo}}(d,\delta,p_{\mathrm{fail}})$, a sufficient number of copies for reconstructing a $d$-dimensional reference state with operator-norm error at most $\delta$ and failure probability at most $p_{\mathrm{fail}}$. This function is supplied by the statistical guarantee of the chosen tomographic procedure and contains all protocol-dependent scaling with the dimension, target accuracy, and confidence level. 

\begin{lemma}
\label{lem:ap_finite_sample}
Consider the task of implementing the Petz map $\mathcal{P}_{\mathcal{N}}^{\sigma}$, associated with a full-rank $d$-dimensional reference state $\sigma$, where $\lambda_{\min}(\sigma)\geq\mu$ and $\lambda_{\min}(\mathcal{N}(\sigma))\geq\nu$, for known constants $\mu,\nu>0$. Let $\hat{\sigma}$ be an estimate of $\sigma$ obtained from $N$ identical copies of $\sigma$, such that 
\begin{equation}
 \Pr\left[\|\sigma-\hat{\sigma}\|_\infty\leq\delta\right]\geq 1-p_{\mathrm{fail}}.
\end{equation}
Define
\begin{equation}
\delta^\star:=\min\left\{\frac{\eta}{C(\mu,\nu,\mathcal N)},\frac{\mu}{2},\frac{\nu}
{2\|\mathcal N(\mathbb 1)\|{\infty}}\right\},
\end{equation}
where
\begin{equation}
    C(\mu, \nu, \mathcal{N}):=\frac{\sqrt{2}\,\|\mathcal{N}(\mathbb{1})\|_\infty}{\nu\sqrt{\mu}}+\frac{4\|\mathcal{N}(\mathbb{1})\|_\infty^2}{\nu^2}.
\end{equation}
Then, to guarantee a target Petz map accuracy $\|
\mathcal{P}_{\mathcal N}^{\sigma}-\mathcal{P}_{\mathcal N}^{\hat{\sigma}}\|_\diamond\leq\eta$ with probability at least $1-p_{\mathrm{fail}}$, while also ensuring that the required spectral condition is satisfied, it is sufficient to require
\begin{equation}
\label{eq:ap_N_star}
N\geq N_{\mathrm{tomo}}\left(
d,\delta^\star,p_{\mathrm{fail}}
\right).
\end{equation}
\end{lemma}
\begin{proof}
By definition, with probability at least $1 - p_{\rm fail}$, 
\begin{equation}
\|\sigma-\hat{\sigma}\|_\infty\leq\delta^\star.
\end{equation}
Since
\begin{equation}
\delta^\star\leq \frac{1}{2}\min\left\{
\mu,\frac{\nu}{\|\mathcal N(\mathbb 1)\|_\infty}\right\},
\end{equation}
the spectral condition required by
Corollary~\ref{cor:bound_spec} is satisfied. Hence, we have 
\begin{equation}
\|\mathcal{P}_{\mathcal N}^{\sigma}-\mathcal{P}_{\mathcal N}^{\hat{\sigma}}\|_\diamond\leq
C(\mu,\nu,\mathcal N)\|\sigma-\hat{\sigma}\|_\infty.
\end{equation}
Finally, because $\delta^\star\leq\eta/C(\mu,\nu,\mathcal N)$, we obtain
\begin{equation}
\|\mathcal{P}_{\mathcal N}^{\sigma}-
\mathcal{P}_{\mathcal N}^{\hat{\sigma}}\|_\diamond
\leq C(\mu,\nu,\mathcal N)\delta^\star
\leq\eta,
\end{equation}
with probability at least $1-p_{\mathrm{fail}}$.\qedhere
\end{proof}
To illustrate how Lemma~\ref{lem:ap_finite_sample} can be combined with a concrete tomographic guarantee, consider the estimation of an arbitrary full-rank state in trace norm. Standard sample-complexity results imply that one may take~\cite{Haah2017}
\begin{equation}
N_{\mathrm{tomo}}
(d,\delta,p_{\mathrm{fail}})= \mathcal{O}\left(\frac{d^2}{\delta^2}\log\frac{1}{p_{\mathrm{fail}}}\right).
\end{equation}
Because $\|\sigma-\hat{\sigma}\|_\infty\leq\|\sigma-\hat{\sigma}\|_1$, this trace-norm guarantee satisfies the hypothesis of the proposition. Substituting
\begin{equation}
\delta^\star=\min\left\{\frac{\eta}{C(\mu,\nu,\mathcal N)},\frac{\mu}{2},\frac{\nu}
{2\|\mathcal N(\mathbb 1)\|_\infty}\right\}
\end{equation}
then gives
\begin{equation}
N=\mathcal{O}\left(d^2\log\frac{1}{p_{\mathrm{fail}}}
\max\left\{\frac{C(\mu,\nu,\mathcal N)^2}{\eta^2},
\frac{4}{\mu^2},\frac{4\|\mathcal N(\mathbb 1)\|_\infty^2}{\nu^2}\right\}\right).
\end{equation}

We now turn to the tomography-assisted implementation of the IPR, illustrated in Fig.~\ref{fig:scheme2}. Because each reference state is estimated from finitely many copies, tomography introduces an additional error at every iteration. Moreover, since the output of one step determines the reference state used at the next, these local errors propagate through the adaptive sequence. In the following proposition, we apply Lemma~\ref{lem:ap_finite_sample} to the successive IPR updates and derive a recursive upper bound on the accumulated implementation error after any finite number of iterations. 

\begin{proposition}[Proof of Proposition~\ref{prop:ipr_error_propagation}]
Consider the exact-reference IPR sequences $\{\rho_k, \sigma_k\}_{k\geq1}$ and their finite-copy implementations $\{\rho_k', \sigma'_k\}_{k\geq1}$. Define the accumulated implementation error $\varepsilon_k:=\|\rho_k'-\rho_k\|_1$. Assume full-rank $d$-dimensional states satisfying
\begin{equation}
    \lambda_{\min}(\sigma_k)\geq\mu_k,\quad \lambda_{\min}(\mathcal{N}(\sigma_k))\geq\nu_k,
\end{equation}
such that at every iteration under consideration,
\begin{equation}
    \delta_k+\frac{\varepsilon_k}{2}\leq\frac{1}{2}\min\left\{\mu_k,\frac{\nu_k}{\left\|\mathcal{N}(\mathbb{1})\right\|_{\infty}}\right\}.
\end{equation}
Here $\mu_k,\nu_k>0$ and $\lambda_{\min}$ denotes the minimal eigenvalue. Then, the accumulated implementation error obeys
\begin{equation}
    \varepsilon_{k+1}\leq\left(1+\frac{C_k}{2}\right)\varepsilon_k+C_k\delta_k\,,
\end{equation}
where 
\begin{equation}
    C_k := \frac{\sqrt{2}\,\|\mathcal{N}(\mathbb{1})\|_\infty}{\nu_k\sqrt{\mu_k}}+\frac{4\|\mathcal{N}(\mathbb{1})\|_\infty^2}{\nu_k^2}.
\end{equation}
\end{proposition}
\begin{proof}
For every $k\geq1$, we have $\sigma'_k=\rho_k'$ and
$\sigma_k=\rho_k$. Therefore,
\begin{align}
\norm{\hat{\sigma}'_k-\sigma_k}_\infty&\leq
\norm{\hat{\sigma}'_k-\sigma'_k}_\infty
+\norm{\sigma'_k-\sigma_k}_\infty \nonumber \\
&\leq\delta_k+\frac{1}{2}\norm{\rho_k'-\rho_k}_1 =\delta_k+\frac{\varepsilon_k}{2}.
\end{align}
Here we used $\norm{A}_\infty\leq\frac12\norm{A}_1$ for a traceless Hermitian operator $A$. Thus, Eq.~\eqref{eq:estimation_spectral_condition} ensures that the spectral condition required by Corollary~\ref{cor:bound_spec}
is satisfied.

We now upper-bound the accumulated error. Using the definitions of the ideal and implemented sequences, we write
\begin{align}
    \rho'_{k+1}\!-\!\rho_{k+1}
    &=\mathcal{P}_{\mathcal{N}}^{\hat{\sigma}'_k}(\rho_k')
    -\mathcal{P}_{\mathcal{N}}^{\sigma_k}(\rho_k)  \\
    &=\mathcal{P}_{\mathcal{N}}^{\hat{\sigma}'_k}(\rho_k'-\rho_k)
    + \left(\mathcal{P}_{\mathcal{N}}^{\hat{\sigma}'_k} -
    \mathcal{P}_{\mathcal{N}}^{\sigma_k}
    \right)(\rho_k). \nonumber
\end{align}
Taking the trace norm and using the triangle inequality gives
\begin{align*}
    \|\rho'_{k+1}\!-\!\rho_{k+1}\|_1&\leq\left\|\mathcal{P}_{\mathcal{N}}^{\hat{\sigma}'_k}
    (\rho_k'\!-\!\rho_k)\right\|_1\!+\!\left\|\left(\mathcal{P}_{\mathcal{N}}^{\hat{\sigma}'_k}\!-\!\mathcal{P}_{\mathcal{N}}^{\sigma_k}\right)(\rho_k)\right\|_1 .
\end{align*}
The first term is bounded by trace-norm contractivity of quantum channels,
\begin{equation}
\left\|\mathcal{P}_{\mathcal{N}}^{\hat{\sigma}'_k}(\rho_k'-\rho_k)\right\|_1\leq \|\rho_k'-\rho_k\|_1 =\varepsilon_k .
\end{equation}
For the second term, we use the definition of the diamond norm:
\begin{equation}
\left\|\left(\mathcal{P}_{\mathcal{N}}^{\hat{\sigma}'_k}-\mathcal{P}_{\mathcal{N}}^{\sigma_k}
    \right)(\rho_k)\right\|_1\leq\left\|\mathcal{P}_{\mathcal{N}}^{\hat{\sigma}'_k}
    -\mathcal{P}_{\mathcal{N}}^{\sigma_k}\right\|_\diamond .
\end{equation}
Using Corollary~\ref{cor:bound_spec}, we obtain 
\begin{equation}
    \left\|\mathcal{P}_{\mathcal{N}}^{\hat{\sigma}'_k}
    -\mathcal{P}_{\mathcal{N}}^{\sigma_k}
    \right\|_\diamond\leq
    C(\mu_k, \nu_k, \mathcal{N})\,\|\hat{\sigma}'_k-\sigma_k\|_\infty \,,
\end{equation}
where 
\begin{equation}
    \mu_k:=\lambda_{\min}(\sigma_k),\quad\text{and}\quad\nu_k:=\lambda_{\min}\!\left(\mathcal{N}(\sigma_k)\right).
\end{equation}
Finally,
\begin{align}
    \|\hat{\sigma}'_k-\sigma_k\|_\infty
    &\leq\|\hat{\sigma}'_k-\sigma'_k\|_\infty + \|\sigma'_k-\sigma_k\|_\infty  \\
    &= \|\hat{\sigma}'_k-\sigma'_k\|_\infty +\|\rho_k'-\rho_k\|_\infty \leq \delta_k+ \frac12\varepsilon_k \nonumber.
\end{align}
In the last step we used the fact that $\sigma'_k-\sigma_k$ is Hermitian and traceless, and hence
\begin{equation}
    \|\sigma'_k-\sigma_k\|_\infty\leq \frac{1}{2}\|\sigma'_k-\sigma_k\|_1.
\end{equation}
We therefore obtain
\begin{equation}
    \varepsilon_{k+1}
    \leq\varepsilon_k
    +C(\mu_k, \nu_k, \mathcal{N})\left(\delta_k+\frac12\varepsilon_k\right)\,.
\end{equation}
\end{proof}
The proposition above should be understood as an error propagation estimate, providing an explicit criterion under which possible tomography-assisted implementations remain controlled.

\begin{theorem}
Let $\psi$ be a pure state and let $\{\rho_k\}_{k\geq1}$ denote the output sequence generated by the exact-reference IPR. For a copy-budget $N$, let $\{\rho_k'\}_{k\geq1}$ denote the output sequence generated by the finite-copy implementation $\mathcal{R}_N$. Suppose that, at iteration $k$, 
\begin{equation}
    \|\rho_k'-\rho_k\|_1\leq \varepsilon_k.
\end{equation}
Then the fidelity achieved by the finite-copy implementation satisfies
\begin{equation}
    F(\psi,\rho_k') \geq F(\psi,\rho_k)-\frac{1}{2} \varepsilon_k\,.
\end{equation}
\end{theorem}
\begin{proof}
Since $\psi$ is pure, the fidelity with any state $\rho$ reduces to $F(\psi,\rho)={\rm Tr}[\psi\rho]$. Therefore,
\begin{align}
    F(\psi,\rho_k')-F(\psi,\rho_k)
    &=
    {\rm Tr}[\psi(\rho_k'-\rho_k)].
\end{align}
Since $\rho_k'-\rho_k$ is Hermitian and traceless, we have
\begin{equation}
    \left|
    {\rm Tr}[\psi(\rho_k'-\rho_k)]
    \right|
    \leq
    \frac{1}{2}
    \|\rho_k'-\rho_k\|_1.
\end{equation}
Hence,
\begin{equation}
    \left|F(\psi,\rho_k')-F(\psi,\rho_k) \right|\leq
    \frac{1}{2}\varepsilon_k.
\end{equation}
In particular,
\begin{equation}
    F(\psi,\rho_k')\geq F(\psi,\rho_k)-\frac{1}{2}\varepsilon_k.
\end{equation}
\end{proof}

\section{Pauli tomography}
\label{ap:pauli}
Let $\sigma = \tfrac12(\mathbb{1} + \vec{r}\cdot\bm\sigma)$ be an unknown qubit state with Bloch vector $\vec{r}\in\mathbb R^3$ where $\|\vec{r}\|_2 \le 1$. Suppose we are given $N$ identical copies of $\sigma$, where $N$ is a multiple of $3$. Let $\hat{\vec R}=(\hat r_x,\hat r_y,\hat r_z)$ be the empirical estimator constructed by measuring $m = N/3$ independent copies in each of the Pauli bases $X, Y$, and $Z$. Let $\hat{\vec r}$ denote the estimated Bloch vector projected back into the Bloch ball as $\hat{\vec r}:=\frac{\vec{R}}{\max\{1,\norm{\vec{R}}_2\}}$. Then, for any target accuracy $\delta > 0$ and a fixed maximum allowed failure probability $p_{\rm fail} \in (0,1)$, to guarantee that the worst-case reconstruction error satisfies 
\begin{equation}
    \Pr\left( \|\sigma - \hat{\sigma}\|_1 >\delta\right)=\Pr\left( \|\hat{\vec r}-\vec r\|_2 >\delta\right) \leq p_{\rm fail} 
\end{equation}
it suffices to choose the total number of copies such that
\begin{equation}
    N \geq \frac{\kappa}{\delta^2} \,, \quad \text{with} \quad \kappa = 18\ln\left(\frac{6}{p_{\rm fail}}\right)\,.
\end{equation}
Fix $m\in\mathbb N$ such that the total number of copies is $N = 3m$. We distribute the copies equally across the three Pauli observables: 
\begin{itemize}
\item Measure $X$ on $m$ independent copies of $\sigma$, obtaining i.i.d.\ outcomes
$X_1,\dots,X_m\in\{+1,-1\}$.
\item Measure $Y$ on $m$ independent copies of $\sigma$, obtaining i.i.d.\ outcomes
$Y_1,\dots,Y_m\in\{+1,-1\}$.
\item Measure $Z$ on $m$ independent copies of $\sigma$, obtaining i.i.d.\ outcomes
$Z_1,\dots,Z_m\in\{+1,-1\}$.
\end{itemize}
Define the empirical vector $\hat{\vec R}=(\hat r_x,\hat r_y,\hat r_z)$ with
\begin{align}
    \hat r_x=\frac1m\sum_{j=1}^m X_j,\,\hat r_y=\frac1m\sum_{j=1}^m Y_j,\,\hat r_z=\frac1m\sum_{j=1}^m Z_j.
\end{align}
Note that $|\hat r_i|\leq 1 \, \forall i\in\{x,y,z\}$ but this empirical vector is not necessarily contained in the Bloch ball. Therefore, define an estimated Bloch vector by projecting $\hat{\vec R}$ into the Bloch ball (if required), i.e., $\hat{\vec r}:=\frac{\vec{R}}{\max\{1,\norm{\vec{R}}_2\}}$. Next, note that
\begin{align}\label{eq:projected_vec_bound}
    \norm{\vec{R}-\vec{r}}_2^2&=\norm{\left(\vec{R}-\hat{\vec r}\right) +\left(\hat{\vec r}-\vec{r}\right) }_2^2 \nonumber \\
    &=\norm{\vec{R}-\hat{\vec r}}_2^2 +\norm{\hat{\vec r}-\vec{r}}_2^2 +2 \braket{\vec{R}-\hat{\vec r}, \hat{\vec r}-\vec{r}} \nonumber \\
    &\geq \norm{\hat{\vec r}-\vec{r}}_2^2,
\end{align}
where the last line follows from~\cite[Theorem~3.16]{Bauschke2017}. Define $\vec{\Delta}=(\Delta_x,\Delta_y,\Delta_z)$ with
\begin{align}
    \Delta_i &:= \hat r_i-r_i \quad \forall i\in \{x,y,z\},
\end{align}
then by Eq.~\eqref{eq:projected_vec_bound} we have $ ||\hat{\vec r}-\vec{r}||_2\leq ||\vec{R}-\vec{r}||_2 =||\vec{\Delta}||_2$. Therefore, for $\delta>0$ we have
\begin{align}\label{eq:union}
    \!\{||\hat{\vec r}-\vec{r}||_2 \!\geq\! \delta\} \!\subseteq\!   \{||\hat{\vec{\Delta}}||_2\!\geq\! \delta\} \!\subseteq  \mkern-7mu \bigcup_{i\in \{x,y,z\}} \{|\Delta_i|\!\geq\! \tfrac{\delta}{\sqrt{3}}\},
\end{align}
where the second inclusion follows because, if all coordinates satisfy $|\Delta_i|<\tfrac{\delta}{\sqrt{3}}$, then $||\vec\Delta ||_2^2=\Delta_x^2+\Delta_y^2+\Delta_z^2<\delta^2$. For each coordinate, the empirical mean is constructed from $m$ independent measurement outcomes in $\{-1,+1\}$, with expectation $r_i$. Hoeffding’s inequality therefore gives, for every $a>0$,
\begin{align}
    \Pr\!\left(|\Delta_i|\ge a\right) \le 2\exp(-\tfrac{ma^2}{2}).
\quad i\in\{x,y,z\}.
\end{align}
Applying the bound in Eq.~\eqref{eq:union} yields
\begin{align}
\Pr\!\left(\| \hat{\vec{r}}-\vec{r}\|_2\ge \delta\right) \!&\leq \! \Pr\!\left(\|\Delta\|_2\ge \delta\right)  \\
&\!\leq\!\sum_{i\in \{x,y,z\}} \Pr\!\left(|\Delta_i|\ge \tfrac{\delta}{\sqrt{3}}\right) \nonumber   \\
&\!\leq \!6 \exp\left(\!-\tfrac{m\delta^2}{6}\right)\!=\!6 \exp\left(\!-\tfrac{N\delta^2}{18}\right). \nonumber
\end{align}
Lastly, note that $\|\sigma-\hat{\sigma}\|_1=\| \hat{\vec{r}}-\vec{r}\|_2$, and thus 
\begin{align}
    \Pr\!\left(\|\sigma-\hat{\sigma}\|_1\ge \delta\right) \leq 6 \exp\left(\!-\tfrac{N\delta^2}{18}\right).
\end{align}
Therefore, to guarantee $\Pr\!\left(\|\sigma-\hat{\sigma}\|_1\ge \delta\right)\leq p_\text{fail}$ it suffices to choose 
\begin{align}
    N\geq \frac{18}{\delta^2}\ln\left(\frac{6}{p_\text{fail}}\right).
\end{align}
Equivalently, for a fixed copy budget $N$, with a probability of at least $1-p_{\text{fail}}$, we have
\begin{align}
    \|\sigma-\hat\sigma\|_1 \leq \sqrt{\frac{18}{N} \ln\left(\frac{6}{p_\text{fail}}\right)}.
\end{align}

\subsection{Tomographic estimates for Petz recovery}
\label{ap:petz_tomography}

The Pauli tomography procedure described above produces an estimated state $\hat{\sigma}$ from $\sigma$, which satisfies 
\begin{equation}
\Pr\!\left(\|\sigma-\hat{\sigma}\|_1\leq\delta_{\mathrm{tomo}}\right)
\geq 1-p_{\mathrm{fail}}, 
\label{eq:tomography_success_event}
\end{equation}
for a given tomographic accuracy $\delta_{\rm tomo}$ and success probability $1-p_{\mathrm{fail}}$. This statement is independent of the subsequent use of the estimate. When $\hat{\sigma}$ is employed as the reference state of a Petz map, however, an additional issue may arise: the inverse $\mathcal N(\hat{\sigma})^{-1/2}$ requires $\mathcal N(\hat{\sigma})$ to be full rank. For example, a pure
estimate aligned with the invariant axis of a single-axis Pauli channel has a rank-deficient channel image. We must therefore treat this case separately in the implementation of the protocol. To this end, we introduce some small perturbation $\epsilon>0$, and use the estimated state 
\begin{equation}
\hat{\sigma}_{\mathcal{P}}= \begin{cases}\hat{\sigma},
& \text{if} \quad \mathcal N(\hat{\sigma})>0,\\[1mm]
(1-\epsilon)\hat{\sigma} +\epsilon\dfrac{\mathbb 1}{2},
&  \text{otherwise},
\end{cases}
\label{eq:petz_compatible_estimate}
\end{equation}
which is well-defined as a reference state for the Petz map. In the perturbed branch,
\begin{equation}
\mathcal N(\hat{\sigma}_{\mathcal{P}})=(1-\epsilon)\mathcal N(\hat{\sigma})+\epsilon\mathcal N(\mathbb 1/2)\geq
\epsilon\mathcal N(\mathbb 1/2).
\end{equation}
Whenever $\mN$ is unital, this yields a full rank state $\mathcal N(\hat{\sigma}_{\mathcal{P}})$. In particular, this is the case for the depolarizing and single-axis Pauli channels considered here. The perturbation in Eq.~\eqref{eq:petz_compatible_estimate} modifies the original tomographic estimate by at most
\begin{equation}
\|\hat{\sigma}_{\mathcal{P}}-\hat{\sigma}\|_1\leq\epsilon.
\end{equation}
Consequently, Eq.~\eqref{eq:tomography_success_event} implies the
general bound
\begin{equation}
\Pr\!\left(\|\sigma-\hat{\sigma}_{\mathcal{P}}\|_1\leq\delta_{\mathrm{tomo}}+\epsilon\right)\geq 1-p_{\mathrm{fail}}.
\label{eq:perturbed_tomography_bound}
\end{equation}
Therefore, if the perturbation error $\epsilon$ is to be included explicitly in a prescribed total estimation accuracy $\delta$, one may choose $\delta_{\mathrm{tomo}}+\epsilon\leq\delta$.

To apply the perturbation bound derived in Appendix~\ref{ap:implement}, assume that
\begin{equation}
    \lambda_{\min}(\sigma)\geq\mu, \qquad
    \lambda_{\min}(\mathcal N(\sigma))\geq\nu .
\end{equation}
If
\begin{equation}
\delta_{\rm tom}\leq\min\left\{\frac{\mu}{2},\frac{\nu}{2\|\mathcal N(\mathbb 1)\|_\infty}\right\},
\end{equation}
then, on the tomography-success event
\(\|\sigma-\hat\sigma\|_1\leq\delta_{\rm tom}\), we have
\begin{equation}
\lambda_{\min}(\hat\sigma)\geq\frac{\mu}{2},\qquad
\lambda_{\min}(\mathcal N(\hat\sigma))\geq\frac{\nu}{2}.
\end{equation}
Consequently, under these spectral assumptions, the perturbation is inactive on the tomography-success event. It may be triggered only on the complementary event, whose probability is at most $p_{\mathrm{fail}}$. The finite-sample perturbation bound therefore remains unchanged.

\section{Finite-copy recovery benchmarks}
\label{ap:comparison}

This section provides additional numerical results to support the analysis in the main text. We examine the performance of the finite-copy implementation of the IPR as a function of copy allocation across the tomographic stages. Furthermore, we compare this multi-copy strategy with single-copy estimation for the depolarizing channel and evaluate measure-and-prepare schemes for the dephasing channel. Because a fixed copy budget induces a probability distribution over measurement outcomes and estimates rather than a deterministic error, all depicted fidelities are averaged over multiple independent tomographic realizations.

\subsection{Copy allocation in the IPR protocol}

The proposed implementation of the IPR protocol employs tomography to estimate the adaptive reference states, as illustrated in Fig.~\ref{fig:scheme2}. Suppose that the protocol is carried out for a total of $K$ iterations. For a fixed copy budget $N$, one must determine how the available copies are distributed among the $K-1$ tomographic stages. The total tomographic budget satisfies
\begin{equation}
    N=\sum_{k=1}^{K-1}N_k\,,
\end{equation}
where $N_k$ denotes the number of copies assigned to the $k$th tomographic stage. We define a copy-allocation scheme as
\begin{equation}
    \pi_j=\left\{N_k^{(j)}\right\}_{k=1}^{K-1},
    \qquad
    \sum_{k=1}^{K-1}N_k^{(j)}=N\,,
\end{equation}
where the index $j$ labels different distributions of the fixed total budget across the tomographic stages. Since the accuracy of each reference-state estimate depends on the number of copies available at the corresponding stage, the allocation can substantially affect the final recovery fidelity. For a fixed total tomographic budget, one may therefore seek the allocation that maximizes the final recovery performance. Conversely, for a prescribed target fidelity and confidence level, one may optimize the allocation to minimize the total number of required copies.

\begin{figure}[h!]
    \centering
    \includegraphics[width=0.8\linewidth]{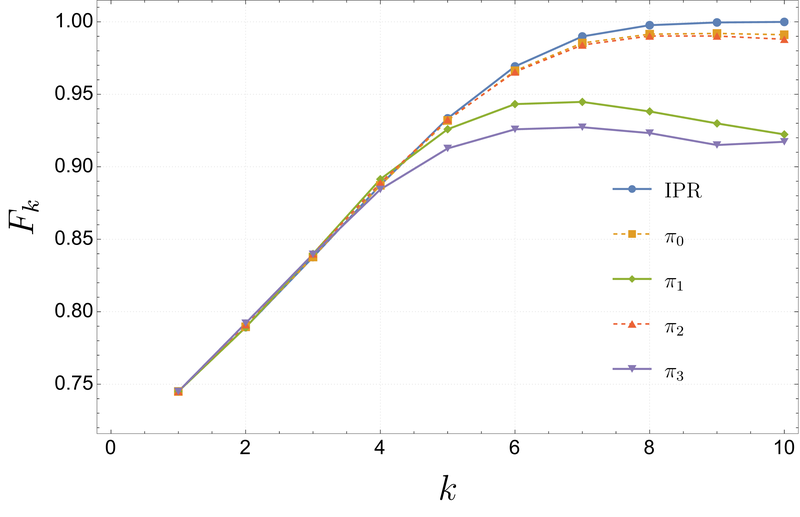}\\
    \includegraphics[width=0.8\linewidth]{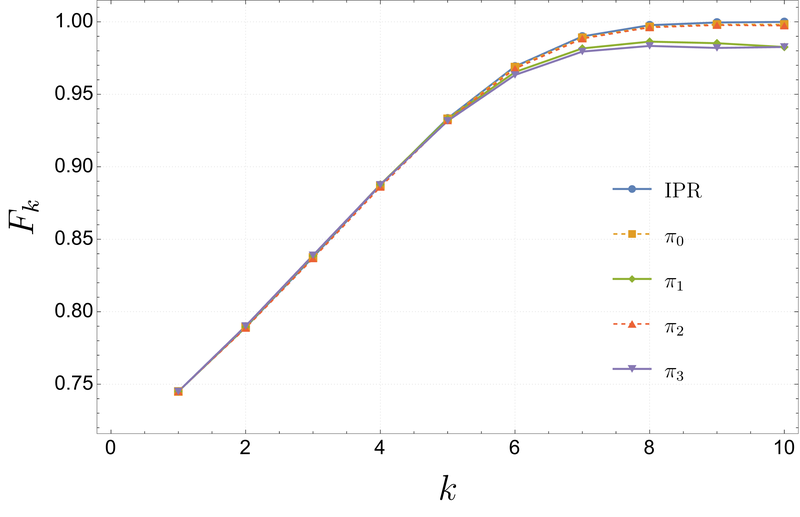}
    \caption{Recovery fidelity of the initial state $\psi_+$ for the depolarizing channel with $p = 0.3$ as a function of the number of iterations. The two panels depict four different copy allocations given a fixed total tomographic copy budget among intermediate stages. The allocations are: uniform $\pi_{0}$; most copies assigned to the initial stages $\pi_{\rm 1}$; symmetric center-weighted allocation with most copies assigned to the intermediate stages $\pi_{\rm 2}$; most copies assigned to the last stages $\pi_{\rm 3}$. Upper panel: Total of $N = 10^4$ copies. Lower panel: Total of $N = 5\times 10^4$ copies. Each point is averaged over $2\times 10^3$ independent tomographic realizations.}
    \label{fig:copies_split}
\end{figure}

Figure~\ref{fig:copies_split} illustrates the influence of different allocation schemes on the fidelity across iterations for the depolarizing channel. We consider four representative schemes: $\pi_0$, in which the copies are distributed uniformly across the tomographic stages; $\pi_1$, in which most copies are assigned to the initial stages; $\pi_2$, a symmetric center-weighted allocation that assigns most copies to the intermediate stages; and $\pi_3$, the reverse of $\pi_1$, in which most copies are assigned to the final stages. For $N=10^4$, shown in the upper panel, the uniform and center-weighted allocations, $\pi_0$ and $\pi_2$, yield the best performance and remain closest to the ideal benchmark. By contrast, the asymmetric allocations $\pi_1$ and $\pi_3$ exhibit a larger degradation in fidelity at later iterations. Increasing the total tomographic budget to $N=5\times10^4$ improves the recovery fidelity for all four schemes, as shown in the lower panel. Nevertheless, the symmetric allocations $\pi_0$ and $\pi_2$ continue to outperform the strongly asymmetric alternatives.  

These results indicate that the total number of copies alone does not fully characterize the finite-resource performance of the IPR protocol: how those copies are distributed among the tomographic stages also plays an important role. In particular, strongly concentrating the available copies at either the beginning or the end of the protocol can create poorly estimated intermediate reference states and thereby limit the overall recovery performance. Although the optimal allocation may depend on the channel, the iteration depth, and the target state, the uniform allocation provides a simple and robust choice when no further optimization is performed.

\subsection{Multi-copy versus single-copy recoverability}

For a Haar-distributed ensemble of pure qubit inputs subject to a depolarizing channel, the optimal fixed, state-independent CPTP post-processing acting on a single noisy copy is the identity channel $\mathcal{I}$. The optimal single-copy fidelity is therefore
\begin{equation}
    \int d\psi\,
    F\!\left(\psi,(\mathcal{I}\circ\mathcal{D}_p)(\psi)\right)
    =1-\frac{p}{2}.
\end{equation}
In fact, owing to the covariance of the depolarizing channel, the integrand itself is independent of $\psi$.

The finite-copy IPR is not subject to the same resource restriction: it uses additional noisy copies to estimate successive reference states and adapt the recovery map. Any improvement over the single-copy benchmark must therefore be interpreted as a multi-copy advantage rather than as a comparison under identical resources. To investigate when such an improvement becomes possible, we consider a uniform allocation in which the same number $N_j$ of copies is assigned to every tomographic stage. The cumulative number of copies consumed before obtaining the output at iteration $k$ is
\begin{equation}
    N_{\mathrm{cumu}}(k)=\sum_{j=1}^{k-1}N_j\,.
\end{equation}

\begin{figure}[ht]
    \centering
    \includegraphics[width=0.8\linewidth]{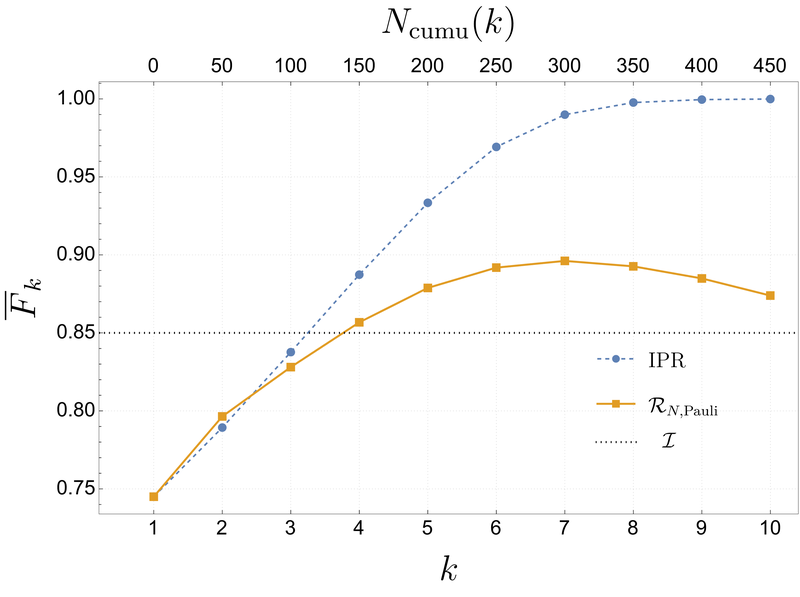}\\
    \includegraphics[width=0.8\linewidth]{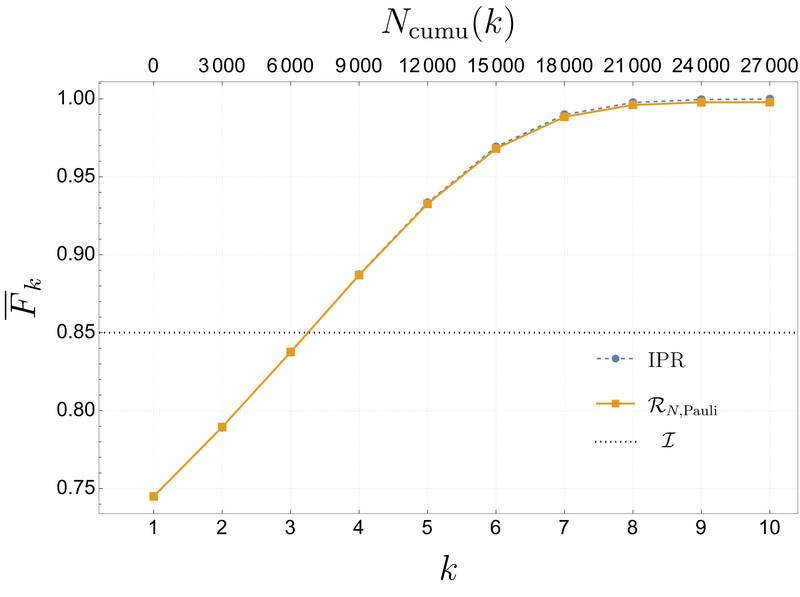}
    \caption{Haar-averaged recovery fidelity as a function of the IPR iteration for the depolarizing channel with $p=0.3$. The finite-copy implementations use $N_j=50$ copies per tomographic stage in the upper panel and $N_j=3000$ in the lower panel. The lower horizontal axis indicates $k$, while the upper axis reports the cumulative number of copies $N_{\mathrm{cumu}}(k)=\sum_{j=1}^{k-1}N_j$. The optimal, state-independent, single-copy benchmark is the identity recovery $\mathcal{I}$ which retains the noisy output, and yields a fidelity $\overline{F} = 1-p/2=0.85$. The ideal IPR result is included as a reference. The finite-copy results are averaged over the Haar-distributed input ensemble and independent tomographic trajectories.
    }
    \label{fig:strategy_single}
\end{figure}

Figure~\ref{fig:strategy_single} compares the resulting finite-copy IPR fidelity with the optimal single-copy benchmark. In the upper panel, $N_j=50$ copies are assigned to each of the $K-1=9$ tomographic stages, corresponding to a total budget of $450$ copies at $K=10$. In this regime, the tomographic errors are sufficiently large that the fidelity need not increase monotonically with the iteration depth. Thus, a larger cumulative number of consumed copies does not necessarily imply better recovery: each additional iteration introduces a new estimation error, while inaccuracies inherited from earlier stages continue to propagate through the adaptive sequence.

In the lower panel, increasing the allocation to $N_j=3000$ copies per stage substantially suppresses these errors. The finite-copy implementation then follows the exact-reference IPR more closely and surpasses the optimal single-copy benchmark after sufficiently many iterations. At small iteration depths, including the first Petz recovery step, the benchmark is not yet exceeded because the reference state has not been refined sufficiently. These results illustrate that the ability of the finite-copy IPR to outperform a single-copy strategy depends jointly on the iteration depth and the tomographic accuracy at each stage. Therefore, reaching a prescribed target fidelity requires controlling the distribution of copies across the protocol, rather than considering only the total cumulative budget.

\subsection{MP for dephasing channel}

In the main text we provided a comparison between the finite-copy IPR and a measure-and-prepare (MP) strategy for the depolarizing channel. We now turn our attention to the dephasing channel given by 
\begin{equation}
\label{eq:dephasing_channel}
    \mathcal{E}_{z,p}(\rho)=(1-p)\rho+p\,\sigma_z\rho\sigma_z,
\end{equation}
for $0\leq p<1/2$. The MP strategy consists of first performing tomography using all available copies of the state, classically inverting it, followed by a projection to the closest state, and then preparing it. An $N$-copy MP protocol is given by 
\begin{equation}
\mathcal{M}_N(\rho^{\otimes N}) =\sum_m {\rm Tr}\left[M_m\rho^{\otimes N}\right]\tau_m ,
\end{equation}
where $\rho = \mathcal{N}(\psi)$ is the noisy state, $\{M_m\}_m$ is a POVM on the $N$-copy input space, and $\tau_m$ is the state prepared upon obtaining outcome $m$.  

For the depolarizing channel, the optimal deterministic measure-and-prepare benchmark can be obtained using the method of~\cite{Gendra2012}. The derivation exploits the full rotational covariance of the noisy input ensemble. In particular, the
depolarized states have Bloch vectors of fixed length and form a single orbit under the action of $\mathrm{SU}(2)$. The measurement can therefore be chosen covariantly, substantially simplifying the
optimization over collective measurements and guessing rules. This construction does not directly extend to dephasing noise. For
a pure input, the dephasing channel produces an output whose Bloch's vector length 
depends on the orientation of the input relative to the dephasing axis. Consequently, the optimal construction of~\cite{Gendra2012} does not provide an optimal benchmark for this channel. Deriving such a benchmark would require a separate optimization over axially covariant collective measurements and their corresponding guessing rules.

\begin{figure}[h!]
    \centering
    \includegraphics[width=\linewidth]{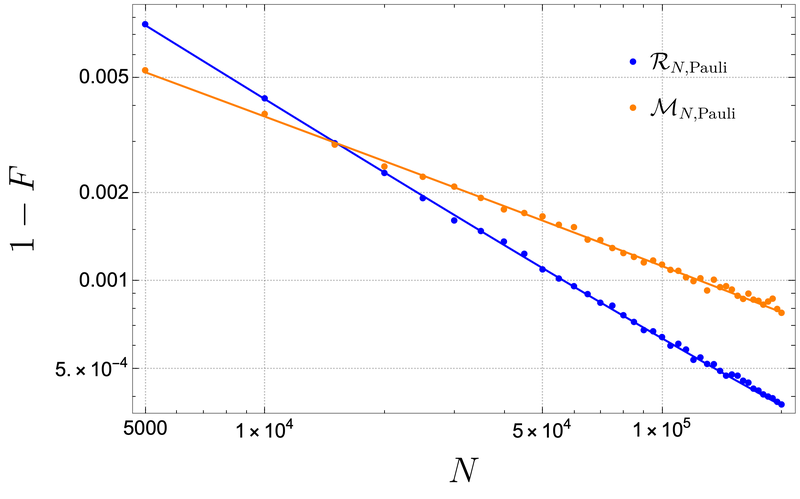}
    \caption{Finite-copy comparison of the implemented iterative Petz recovery protocol, IPR$(N)$, with the MP using Pauli tomography. The infidelity, $1-F$, after a total of $K = 10$ IPR steps is plotted against the total tomographic copy budget $N$ on logarithmic axes for a dephasing channel
    with $p=0.15$ and input state $|\psi\rangle=|+\rangle$. The finite-copy IPR fidelity is averaged over $2\times10^{3}$ independent tomographic trajectories.
    }
    \label{fig:comp_dephasing}
\end{figure}

Figure~\ref{fig:comp_dephasing} therefore compares the finite-copy IPR with the specified Pauli tomographic measure-and-prepare strategy $\mathcal{M}_{N, \rm Pauli}$. The observed crossover with the IPR occurs at a smaller copy number under dephasing noise than under depolarizing noise. The favorable geometry of the depolarizing channel explains the high fidelities achievable by the MP strategy using relatively fewer copies than the IPR. This comparison clarifies the different operational restrictions underlying the two approaches. A meaningful potential advantage of the IPR should be sought in settings in which the recovered physical system must be retained, correlations must be preserved, or fresh state preparation is not permitted. Less symmetric channels also provide an important test, since their inverse action is no longer reduced to a simple radial rescaling of the Bloch vector.

 \makeatletter
\renewcommand{\bibsection}{%
  \par
  \begingroup
    \baselineskip=26pt
    \bib@device{\columnwidth}{245.5pt}%
  \endgroup
  \nobreak\@nobreaktrue
  \addvspace{19pt}%
  \par
}
\makeatother

\bibliography{references}

\end{document}